\documentclass[useAMS]{gGAF2e}
\usepackage{amssymb}

\newcommand{\bbb}{{\bf b}}
\newcommand{\bB}{{\bf B}}

\newcommand{\be}{{\bf e}}

\newcommand{\bscE}{{\boldmath \cal{E}}}
\newcommand{\bF}{{\bf F}}
\newcommand{\bscF}{{\boldmath \cal{F}}}
\newcommand{\bff}{{\bf f}}
\newcommand{\bg}{{\bf g}}
\newcommand{\bG}{{\bf G}}

\newcommand{\bI}{{\bf I}}

\newcommand{\bJ}{{\bf J}}
\newcommand{\bk}{{\bf k}}
\newcommand{\bK}{{\bf K}}
\newcommand{\bL}{{\bf L}}

\newcommand{\bR}{{\bf R}}
\newcommand{\bT}{{\bf T}}
\newcommand{\bu}{{\bf u}}
\newcommand{\bU}{{\bf U}}
\newcommand{\bv}{{\bf v}}
\newcommand{\bw}{{\bf w}}
\newcommand{\bW}{{\bf W}}
\newcommand{\bx}{{\bf x}}

\newcommand{\bY}{{\bf Y}}

\newcommand{\bZ}{{\bf Z}}

\newcommand{\bxi}{{\bm \xi}}

\newcommand{\bpsi}{{\bm \psi}}
\newcommand{\bchi}{{\bm \chi}}
\newcommand{\bOmega}{{\bm \Omega}}

\newcommand{\bmB}{\overline{{\bf B}}}
\newcommand{\bmF}{\overline{{\bf F}}}
\newcommand{\bmU}{\overline{{\bf U}}}
\newcommand{\bnab}{\nabla}
\newcommand{\cE}{\cal{E}}
\def\bzo {{\bf 0}}
\def\mB {\overline{B}}
\def\mU {\overline{U}}
\def\dd {\mbox{d}}
\def\hb {\hat{b}}
\def\hu {\hat{u}}
\def\iu {\mbox{i}}
\def\x {\times}
\def\ol {\overline}
\def\p {\partial}

\begin{document}

\doi{10.1080/03091920xxxxxxxxx} \issn{1029-0419} \issnp{0309-1929}
 \jvol{00} \jnum{00} \jyear{2006} \jmonth{XXXX}

\markboth{Karl-Heinz R\"{a}dler \& Matthias Rheinhardt}
{Mean--field electrodynamics: Critical analysis of various analytical approaches}

\title{Mean--field electrodynamics:\\
Critical analysis of various analytical approaches to the mean electromotive force}
\author{KARL-HEINZ R\"{A}DLER * and MATTHIAS RHEINHARDT \\
Astrophysikalisches Institut Potsdam, An der Sternwarte 16, D-14482 Potsdam, Germany \\
\thanks{ \vspace{6pt}
\newline{\tiny{ {\em } * Corresponding author. E-mail khraedler@arcor.de }}}}
\received{Received 10th April, 2006; in final form 6th November, 2006}

\maketitle

\begin{abstract}
There are various analytical approaches to the mean electromotive force $\bscE = \langle \bu \x \bbb \rangle$
crucial in mean--field electrodynamics, with $\bu$ and $\bbb$ being velocity and magnetic field fluctuations.
In most cases the traditional approach, restricted to the second--order correlation approximation,
has been used.
Its validity is only guaranteed for a range of conditions,
which is narrow in view of many applications, e.g., in astrophysics.
With the intention to have a wider range of applicability other approaches have been proposed
which make use of the so--called $\tau$--approximation, reducing correlations of third order
in $\bu$ and $\bbb$ to such of second order.
After explaining some basic features of the traditional approach a critical analysis of the approaches
of that kind is given.
It is shown that they lead in some cases to results which are in clear conflict
with those of the traditional approach.
It is argued that this indicates shortcomings of the $\tau$--approaches
and poses serious restrictions to their applicability.
These shortcomings do not result from the basic assumption of the $\tau$--approximation.
Instead, they seem to originate in some simplifications made in order to derive $\bscE$
without really solving the equations governing $\bu$ and $\bbb$.
A starting point for a new approach is described which avoids the conflict.
\end{abstract}


\section{Introduction}

In mean--field electrodynamics the mean electromotive force
$\bscE = \langle \bu \x \bbb \rangle$ due to the fluctuations $\bu$ and $\bbb$
of the fluid velocity and the magnetic field plays a crucial role
\citep{krauseetal71b,krauseetal80,moffatt78,raedler00b}.
A central problem is the determination of $\bscE$ for a given motion
as a functional of the mean magnetic field.
Various methods have been used for that.

One approach, which we call ``traditional approach"
or ``approach (i)" in the following, was established together
with mean--field electrodynamics at all \citep{krauseetal71b,krauseetal80}.
Most of the calculations of $\bscE$ have been done on this basis
using the so--called second--order correlation approximation (SOCA)
or, what means the same, first--order smoothing approximation (FOSA).
This approximation in its original form, that is, applied in the case
of a purely hydrodynamic background turbulence, ignores all higher
than second--order correlations in the fluctuations $\bu$ of the velocity field
(see section~\ref{subsec31}).
It can be justified only in cases in which these fluctuations are not too large.
The usually given simple sufficient conditions for its validity are
in view of astrophysical applications rather narrow.
Basically it is possible to proceed to higher--order approximations
but this requires tremendous efforts and has been done so far
only in a few simple cases
(Nicklaus and Stix 1988, Carvalho 1992, 1994, R\"adler {\it et al.} 1997,
see also section~\ref{new1}).
A slight modification of the second--order correlation approximation applies also
to the case of a magnetohydrodynamic background turbulence
(see section~\ref{subsec32}).

In some more recent investigations (which are cited below) other approaches are used,
which rely in a sense on the $\tau$--approximation of turbulence theory \citep{orszag70}
and are called ``$\tau$-approaches" or ``approaches (ii)" in the following.
They go in so far beyond the second--order correlation approximation
as they consider also higher--order correlations,
which are then in the sense of a closure expressed by second--order ones.
In approach (i) the relevant equations are simplified by a well--defined approximation
and then really solved, and $\bscE$ is calculated with these solutions.
In the approaches (ii) a relation for $\bscE$ is deduced from the original equations,
but without really solving them.
Instead, in order to get manageable results, assumptions on the connection
of some of the occurring quantities with  $\bscE$ are introduced.
The final result for $\bscE$ is to a large extent determined by these assumptions.
The approaches (ii) cover from the very beginning also the case of a magnetohydrodynamic
background turbulence.

The main purpose of this paper is a critical analysis of the approaches (ii).
Each step of approach (i) can be justified by the underlying induction equation
or, in the case of a magnetohydrodynamic background turbulence, induction equation
and momentum balance.
Clear, at least sufficient conditions for the applicability of the second--order approximation
can be given.
There is no doubt in the correctness of its results in the so defined range of applicability.
It seems reasonable to assume, and we do so in this paper, that there is at least some overlap
of the ranges of applicability of the approaches (i) and (ii).
We have then to require that in these overlapping ranges the results of both approaches coincide.
Simple versions of approach (ii) as used by \citet{vainshteinetal83},
\citet{blackmanetal02b} and \cite{brandenburgetal05b},
called ``simple $\tau$--approach" or ``approach (iia)" in the following, deliver results which do not
in all cases satisfy this requirement.
As we will show below the more sophisticated version used in the papers
by \cite{raedleretal03} and by \citet{rogachevskiietal03,rogachevskiietal04},
called ``spectral $\tau$--approach" or ``approach (iib)", does not satisfy this requirement, too.
We have to conclude that these approaches are not in full accordance with the basic equations mentioned.
Therefore the results can not be taken for granted.
We will propose a starting point for an alternative approach which avoids conflicts with approach (i).

In section~\ref{model} we define the frame of our considerations
and deliver the basic equations.
In section~\ref{seci} we recall the fundamentals of approach (i) and review some of its basic results.
In section~\ref{secii} we explain the approaches (iia) and (iib),
derive a few results, restricting attention to the simple case of a non--rotating fluid,
and pinpoint shortcomings of these approaches and deviations of the results from those of approach (i).
In section~\ref{new} we explain a proposal for the alternative approach mentioned.
Finally in Section~\ref{sum} we summarize our findings.

\section{The mean--field concept in magnetofluiddynamics}
\label{model}

Let us first define the frame of our considerations.
We want to explain essential features of the various approaches to the mean electromotive force $\bscE$
but do not strive to a high level of generality.

\subsection{Mean--field electrodynamics}

Let us consider the magnetic field $\bB$ in an electrically conducting moving fluid.
We assume that it is governed by the induction equation
\begin{equation}
\p_t \bB = \eta \nabla^2 \bB
    + \bnab \x (\bU \x \bB) \, , \quad
    \bnab \cdot \bB = 0 \, .
\label{eq101}
\end{equation}
Here $\bU$ means the velocity of the fluid and $\eta$ its magnetic diffusivity, assumed to be constant.
Until further notice we consider the fluid motion, that is $\bU$, as given.

We assume that the motion and therefore also the magnetic field shows irregular, e.g. turbulent,
features.
Any field $F$ of this type is split into a mean field $\ol{F}$ and a ``fluctuating" field $f$,
that is, $F = \ol{F} + f$.
The mean field $\ol{F}$ is defined as an average of $F$.
It is assumed that the averaging procedure satisfies the Reynolds averaging rules.
Alternatively to the notation $\ol{F}$ we use also $\langle F \rangle$ in the following.

Averaging the induction equation (\ref{eq101}) we obtain the mean--field induction equation
\begin{equation}
\p_t \bmB = \eta \nabla^2 \bmB
    + \bnab \x (\bmU \x \bmB + \bscE) \, , \quad
    \bnab \cdot \bmB = 0
\label{eq103}
\end{equation}
with the mean electromotive force $\bscE$ given by
\begin{equation}
\bscE = \ol{\bu \x \bbb} \, .
\label{eq105}
\end{equation}

Elaboration of mean--field electrodynamics means studying the properties of this quantity.
For this purpose we rely on the equation for the fluctuations $\bbb$
which follows from (\ref{eq101}) and (\ref{eq103}),
\begin{equation}
\p_t \bbb = \eta \nabla^2 \bbb
    + \bnab \x (\bmU \x \bbb + \bu \x \bmB)  + \bG \, , \quad
    \bnab \cdot \bbb = 0 \, .
\label{eq107}
\end{equation}
Here $\bG$ stands for a term of second order in $\bu$ and $\bbb$,
\begin{equation}
\bG = \bnab \x (\bu \x \bbb)' \, ,
\label{eq109}
\end{equation}
with $(\bu \x \bbb)' = \bu \x \bbb - \overline{\bu \x \bbb}$.

As can be concluded from (\ref{eq105}) and (\ref{eq107}) the mean electromotive force $\bscE$ depends,
apart from initial and boundary conditions for $\bbb$, on $\bmU$, $\bu$ and $\bmB$.
More precisely, $\bscE$ is a functional of these quantities in the sense that $\bscE$
in a given point in space and time depends on $\bmU$, $\bu$ and $\bmB$ also in other space points
and at past times.
This functional is linear in $\bmB$.
For most of the applications it is reasonable to assume that it is in addition local in the sense
that $\bscE$ in a point in space and time depends only on $\bmU$, $\bu$ and $\bmB$
in a certain surroundings of this point.
Likewise in many applications the variations of $\bmB$ in space and time are sufficiently weak
so that the behavior of $\bmB$ inside the relevant surroundings of a given point
can be represented by $\bmB$ and its first spatial derivatives in this point.
Then we have
\begin{equation}
{\cE}_i = {\cE}_i^{(0)} + a_{ij} \mB_j + b_{ijk} \frac{\p \mB_j}{\p x_k} \, ,
\label{eq111}
\end{equation}
where the quantities ${\cE}_i^{(0)}$, $a_{ij}$ and $b_{ijk}$ are functionals of $\bmU$ and $\bu$
in the above sense but do not depend on $\bmB$.

Assume for a simple example that there is no mean motion, $\bmU = \bzo$, and that $\bu$ corresponds
to a homogeneous isotropic, not necessarily mirror--symmetric turbulence.
Since no isotropic vector exists, ${\cE}_i^{(0)}$ must be equal to zero.
Further $a_{ij}$ and $b_{ijk}$ have to be isotropic tensors,
$a_{ij} = \alpha \, \delta_{ij}$ and $b_{ijk} = \beta \, \epsilon_{ijk}$,
with factors $\alpha$ and $\beta$ being averaged quantities depending on $\bu$.
Consequently we have
\begin{equation}
\bscE = \alpha \, \bmB - \beta \, \bnab \x \bmB \, .
\label{eq113}
\end{equation}
For another simple example we remain with $\bmU = \bzo$ and assume that $\bu$ corresponds
to an inhomogeneous turbulence in a rotating system, that is, under the influence of a Coriolis force.
More precisely, $\bu$ deviates from a homogeneous isotropic mirror--symmetric turbulence
only by an inhomogeneity and therefore an anisotropy described by a vector $\bg$ parallel
to the intensity gradient, and by the anisotropy and the deviation from mirror--symmetry
described by the angular velocity $\bOmega$ responsible for the Coriolis force.
Then ${\cE}_i^{(0)}$ needs no longer to be zero,
and $a_{ij}$ and $b_{ijk}$ are no longer purely isotropic tensors.
For the sake of simplicity we nevertheless at first ignore ${\cE}_i^{(0)}$, referring to the comments
in sections \ref{seci}, \ref{subsubsec311}, \ref{subsubsec314}, \ref{subsubsec321} and \ref{new1}.
We further assume that the influences of $\bg$ and $\bOmega$ on the turbulence
are weak enough so that $\bscE$ is linear in these quantities.
Considering then the symmetry properties of the basic equations
(see, e.g., Krause and R\"adler 1971, 1980)
we have
\begin{eqnarray}
\bscE &=& - \gamma \, \bg \x \bmB
   - \alpha_1 \, (\bg \cdot \bOmega) \bmB
   - \alpha_2 \, (\bg \cdot \bmB)\bOmega - \alpha_3 \, (\bOmega \cdot \bmB)\bg
\nonumber\\
&& - \beta \, \bnab \x \bmB
   - \delta_1 \, (\bOmega \cdot \bnab) \, \bmB - \delta_2 \, \bnab (\bOmega \cdot \bmB) \, ,
\label{eq115}
\end{eqnarray}
where the coefficients $\gamma$, $\alpha_1$, $\ldots$ $\delta_2$ are determined by $\bu$ only.
(The choice of the signs in (\ref{eq115}), which follows some other papers,
is however unimportant at this place.)

Given the structures of $\bscE$ as in these examples the only remaining task is the calculation
of coefficients like $\alpha$, $\beta$, etc.
The approximations we want to discuss in this paper concern only these calculations.

\subsection{Dynamical aspects}

Later we will relax the assumption that the fluid motion, that is $\bU$, is given.
In preparation to this we provide here the relevant relations.
Restricting our attention to an incompressible fluid we assume that $\bU$ is governed
by the momentum balance,
\begin{equation}
\varrho \, \Big( \p_t \bU + (\bU \cdot \bnab) \bU \Big) = - \bnab P + \varrho \, \nu \, \nabla^2 \bU
    - 2 \varrho \, \bOmega \x \bU
    + \frac{1}{\mu} (\bnab \x \bB) \x \bB + \varrho \, \bF \, , \quad
    \bnab \cdot \bU = 0 \, .
\label{eq121}
\end{equation}
Here $\varrho$ is the mass density, $P$ the hydrodynamic (including centrifugal) pressure,
$\nu$ the kinematic viscosity, considered as constant, and $\bF$ means an external force.
A rotating frame of reference is assumed, with $\bOmega$ being the angular velocity that defines
the Coriolis force.

Taking the average of equations (\ref{eq121}) we obtain
\begin{equation}
\varrho \, \big( \p_t \bmU + (\bmU \cdot \bnab) \bmU \big) = - \bnab \ol{P} + \, \varrho \, \nu \nabla^2 \bmU
    - 2 \varrho \, \bOmega \x \bmU
    + \frac{1}{\mu} (\bnab \x \bmB) \x \bmB + \varrho \,(\bmF + \bscF) \, , \;
    \bnab \cdot \bmU = 0 \, ,
\label{eq123}
\end{equation}
with a mean ponderomotive force $\bscF$ given by
\begin{equation}
\bscF = - \ol{(\bu \cdot \bnab) \bu} + \frac{1}{\mu \varrho}\ol{(\bnab \x \bbb) \x \bbb} \, .
\label{eq125}
\end{equation}

In view of both $\bscE$ and $\bscF$ the equations for the fluctuations $\bu$ are of interest.
They can be concluded from (\ref{eq121}) and (\ref{eq123}),
\begin{eqnarray}
&& \!\!\!\!\!\!\!\!\! \p_t \bu  + (\bmU \cdot \bnab) \bu + (\bu \cdot \bnab) \bmU
\nonumber\\
&& \quad
    = - \frac{1}{\varrho} \, \bnab(p + \frac{1}{\mu} \bmB \cdot \bbb)
    + \nu \nabla^2 \bu
    - 2 \, \bOmega \x \bu
    + \frac{1}{\mu \varrho} \big( (\bmB \cdot \bnab) \bbb + (\bbb \cdot \bnab) \bmB \big)
    + \bff
    + \bT \, , \quad
    \bnab \cdot \bu = 0 \, .
\label{eq127}
\end{eqnarray}
$\bT$ stands for the terms of second order in $\bu$ and $\bbb$,
\begin{equation}
\bT = - \big((\bu \cdot \bnab) \bu \big)' - \frac{1}{2 \mu \varrho} \bnab (\bbb^2)'
    + \frac{1}{\mu \varrho} \big((\bbb \cdot \bnab) \bbb\big)' \, ,
\label{eq129}
\end{equation}
where $\big((\bu \cdot \bnab) \bu \big)' = (\bu \cdot \bnab) \bu  - \langle (\bu \cdot \bnab) \bu \rangle$ etc.

For many purposes it is necessary to eliminate the pressure term
$\frac{1}{\varrho} \, \bnab(p + \frac{1}{\mu} \bmB \cdot \bbb)$ in  (\ref{eq127}).
Assuming an infinitely extended fluid we find
\begin{equation}
\p_t \bu  =  \nu \nabla^2 \bu
    + \tilde{\bR} + \tilde{\bL} + \tilde{\bU}
    + \tilde{\bff} + \tilde{\bT} \, , \quad
    \bnab \cdot \bu = 0 \, ,
\label{eq141}
\end{equation}
with
\begin{eqnarray}
\tilde{\bR} &=& 2 \,(\bOmega \cdot \bnab) \bpsi \, , \quad
    \bpsi (\bx) = \frac{1}{4 \pi} \int_\infty
    \frac{\bnab' \x \bu (\bx')}{|\bx - \bx'|} \, \dd^3 x'
\nonumber\\
\tilde{\bL} &=& \frac{1}{\mu \varrho} \,
    \big((\bmB \cdot \bnab) \, \bbb + (\bbb \cdot \bnab) \, \bmB
    + \bnab M \big) \, , \quad
    M (\bx) = \frac{1}{2 \pi} \int_\infty
    \frac{\p \mB_i (\bx')/ \p {x_j}' \cdot
    \p b_j (\bx')/ \p {x_i}'}{|\bx - \bx'|} \, \dd^3 x'
\nonumber\\
\tilde{\bU} &=& - (\bmU \cdot \bnab) \bu - (\bu \cdot \bnab) \bmU
    - \bnab V \, , \quad
    V (\bx) = \frac{1}{2 \pi} \int_\infty
    \frac{\p \mU_i (\bx')/ \p {x_j}' \cdot
    \p u_j (\bx')/ \p {x_i}'}{|\bx - \bx'|} \, \dd^3 x'
\label{eq143}\\
\tilde{\bff} &=& \bff + \bnab g \, , \quad
    g (\bx) = \frac{1}{4 \pi} \int_\infty
    \frac{\bnab' \cdot \bff (\bx')}{|\bx - \bx'|} \, \dd^3 x'
\nonumber\\
\tilde{\bT} &=& \bT  + \bnab W \, , \quad
    W (\bx) = \frac{1}{4 \pi} \int_\infty
    \frac{\bnab' \cdot \bT (\bx')}{|\bx - \bx'|} \, \dd^3 x' \, .
\nonumber
\end{eqnarray}
The fields $\tilde{\bL}$, $\tilde{\bR}$, $\tilde{\bU}$, $\tilde{\bff}$, and $\tilde{\bT}$
are by construction divergence--free.
The vector potential $\bpsi$ of $\bu$ satisfies
\begin{equation}
\bnab \x \bpsi = \bu \, , \quad  \bnab \cdot \bpsi = 0 \, .
\label{eq144}
\end{equation}

For various purposes it is reasonable to consider $\p \mB_i / \p x_j$
as constant.
Then we have
\begin{equation}
M (\bx) = 2 \, \frac{\p \mB_i}{\p x_j} \,
    \frac{\p {\tilde{b}}_j (\bx)}{\p x_i} \, ,  \quad
    {\tilde{b}}_j (\bx) = \frac{1}{4 \pi} \int_\infty
    \frac{b_j (\bx')}{|\bx - \bx'|} \, \dd^3 x' \, .
\label{eq145}
\end{equation}
We note that $\bnab \cdot \tilde{\bbb} = 0$ and $\nabla^2 \tilde{\bbb} = - \bbb$.
Analogous relations apply to $V$ if $\p \mU_i / \p x_j$ is constant.
All quantities that occur in (\ref{eq143}) to (\ref{eq145}) may depend also on time,
but for the sake of simplicity we have dropped the argument $t$ everywhere.

\section{The traditional approach (approach (i))
to the mean electromotive force $\bscE$ }
\label{seci}

In view of a later comparison with $\tau$--approaches we repeat here basic ideas
of the traditional approach to the mean electromotive force $\bscE$ and summarize some general results
(see also, e.g., Krause and R\"adler 1980 or R\"adler 2000).
We recall that many applications have been made in the astrophysical context
(e.g., \citet{ruedigeretal93b}, \citet{kitchatinovetal94b}).
For the sake of simplicity we exclude any mean motion of the fluid, that is, $\bmU = \bzo$.
If $\bu$ is given, $\bbb$ is determined by equation (\ref{eq107}) and proper initial and boundary conditions.
This equation is inhomogeneous in $\bbb$ due to the term with $\bmB$.
Any solution can be considered as a superposition of a solution of the homogeneous equation,
which is independent of $\bmB$, and a solution of the full equation
which is linear and homogeneous in $\bmB$.
If there is a non--decaying solution of the homogeneous equation
the mean electromotive force $\bscE$ may have a non--decaying part, say $\bscE^{(0)}$,
independent of $\bmB$ (see, e.g., R\"adler 1976, 2000).
If $\bnab \x \bscE^{(0)} = \bzo$ such solutions correspond to small--scale dynamos.
Until further notice we assume however that $\bbb$ decays to zero as soon as $\bmB$ is equal to zero.
In this case, discussed in section \ref{subsec31}, we speak of ``purely hydrodynamic background turbulence".
Later, in section \ref{subsec32}, we will admit a non--decaying $\bbb$ even if $\bmB$ is equal to zero
and then speak of ``magnetohydrodynamic background turbulence".
We use the word ``turbulence" here in a wide sense.
If not specified otherwise $\bu$ and $\bbb$ are considered as fields
with arbitrary space and time dependencies but, of course, with zero averages.

\subsection{Purely hydrodynamic background turbulence}
\label{subsec31}

\subsubsection{Second--order correlation approximation}
\label{subsubsec311}

We start here with  the induction equation (\ref{eq107}) for the magnetic fluctuations $\bbb$.
As already mentioned we exclude any mean motion, that is, put $\bmU = \bzo$.
We use the second--order correlation approximation, here defined by neglecting the term $\bG$
in equation (\ref{eq107}).
Then this equation takes the simple form
\begin{equation}
\p_t \bbb - \eta \nabla^2 \bbb = \bnab \x (\bu \x \bmB) \, .
\label{eq200}
\end{equation}
Until further notice a non--zero divergence of $\bu$ is admitted.
We assume that equation (\ref{eq200}) applies in all infinite space.
Then its general solution can be written as
\begin{eqnarray}
b_k (\bx, t) &=& \int_\infty G^{(\eta)} (\bx - \bx', t - t_0) \, b_k (\bx', t_0) \, \dd^3 x'
\nonumber\\
&& \quad \quad \quad
+ \epsilon_{klm} \epsilon_{mnp} \, \int_\infty \int_{t_0}^t
    G^{(\eta)} (\bx - \bx', t - t') \frac{\p}{\p x'_l}
    \big( u_n (\bx', t') \, \mB_p (\bx', t') \big) \, \dd^3 x' \, \dd t' \, ,
\label{eq201}
\end{eqnarray}
where $G^{(\eta)}$ is a Green's function,
\begin{equation}
G^{(\eta)} (\bx, t) = G^{(\eta)} (x, t) = (4 \pi \eta t)^{-3/2}
    \exp(- x^2 / 4 \eta t) \, ,
\label{eq203}
\end{equation}
and $t = t_0$ means an initial time.
We note that here, as a consequence of the neglect of $\bG$, solutions $\bbb$ which do not decay
for $\bmB = \bzo$ are automatically excluded.

For the calculation of $\bscE$ we start from
\begin{equation}
{\cE}_i (\bx, t) = \epsilon_{ijk} \langle u_j (\bx, t) \, b_k (\bx, t) \rangle \, ,
\label{eq205}
\end{equation}
consider $t - t_0$ as sufficiently large so that $\langle u_j (\bx, t) \, b_k (\bx', t_0) \rangle = 0$
and let, for simplicity only, finally $t_0 \to - \infty$.
In this way we find
\begin{equation}
{\cE}_i (\bx, t) =  \big( \epsilon_{ijn} \delta_{lp} - \epsilon_{ijp} \delta_{ln} \big) \,
     \int_\infty \int_{0}^\infty \frac{\p G^{(\eta)} (\xi, \tau)}{\p \xi}
     \frac{\xi_l}{\xi} \, Q_{jn} (\bx, t; - \bxi, - \tau) \, \mB_p (\bx - \bxi, t - \tau) \, \dd^3 \xi \, \dd \tau
\label{eq207}
\end{equation}
with the second--rank velocity correlation tensor $Q_{ij}$ defined by
\begin{equation}
Q_{ij} (\bx, t; \bxi, \tau) = \langle u_i (\bx, t) \, u_j (\bx + \bxi, t + \tau) \rangle \, .
\label{eq208}
\end{equation}
We further assume sufficiently weak variations of $\bmB$ in $\bx$ and $t$,
and put under the integral in the sense of a truncated Taylor expansion
\begin{equation}
\mB_p (\bx - \bxi, t - \tau) = \mB_p (\bx, t)
     - \xi_q \frac{\p \mB_p (\bx, t)}{\p x_q} \, ,
\label{eq209}
\end{equation}
ignoring all other terms, that is, those with
$\tau \p \mB_p / \p t$,
$\xi_q \xi_r \p^2 \mB_p / \p x_q \p x_r$, $\ldots \,$ .
In this way we find relation (\ref{eq111}) with ${\cE}_i^{(0)} = 0$,
that is ${\cE}_i = a_{ip} \mB_p + b_{ipq} \p \mB_p / \p x_q$, and
\begin{eqnarray}
a_{ip} &=& - \, \big( \epsilon_{ijn} \delta_{lp} - \epsilon_{ijp} \delta_{ln} \big) \,
     \int_\infty \int_{0}^\infty \frac{\p G^{(\eta)} (\xi, \tau)}{\p \xi}
     \frac{\xi_l}{\xi} Q_{jn} (\bx, t; \bxi, - \tau) \, \dd^3 \xi \, \dd \tau
\nonumber\\
&=&  \big(\epsilon_{ijn} \delta_{lp} - \epsilon_{ijp} \delta_{ln}\big)
     \int_\infty \int_{0}^\infty G^{(\eta)} (\xi, \tau)\,
     \frac{\p Q_{jn} (\bx, t; \bxi, - \tau)}{\p \xi_l} \, \dd^3 \xi \, \dd \tau
\label{eq213}\\
b_{ipq} &=& - \, \big( \epsilon_{ijn} \delta_{lp} - \epsilon_{ijp} \delta_{ln} \big) \,
     \int_\infty \int_{0}^\infty \frac{\p G^{(\eta)} (\xi, \tau)}{\p \xi}
     \frac{\xi_l \xi_q}{\xi} Q_{jn} (\bx, t; \bxi, - \tau) \, \dd^3 \xi \, \dd \tau  \, .
\nonumber
\end{eqnarray}

\subsubsection{Range of validity}
\label{subsubsec312}

In order to formulate conditions under which the above results apply
we introduce the magnetic Reynolds number $Rm$ and the Strouhal number $St$,
\begin{equation}
Rm = u_c \, \lambda_c / \eta \, , \quad St = u_c \tau_c / \lambda_c \, .
\label{eq220}
\end{equation}
Here $u_c$ means a characteristic value of the fluctuating velocity $\bu$,
and $\lambda_c$ and $\tau_c$ are a characteristic length and a characteristic time
of the variations of $\bu$.
If $\bu$ describes a turbulence $\lambda_c$ and $\tau_c$ can be interpreted as correlation length
and correlation time.
We further define the dimensionless parameter $q$ by
\begin{equation}
q = \lambda_c^2 / \eta \tau_c \, .
\label{eq221}
\end{equation}
It gives the ratio of the magnetic diffusion time $\lambda_c^2 / \eta$
for a fluid element whose size is characterized
by $\lambda_c$ to the time $\tau_c$.
We have then $Rm / St = q$.
If, e.g., $St$ is fixed, the limit $Rm \to 0$ coincides with $q \to 0$,
and $Rm \to \infty$ with $q \to \infty$.
Sometimes the limits $q \to \infty$ and $q \to 0$ are denoted by ``high--conductivity limit"
and ``low--conductivity limit", respectively.
This notation, introduced with the idea of varying $\eta$ for fixed finite values
of $\lambda_c$ and $\tau_c$, might however be misleading.
If, e.g., $\lambda_c$ and $\eta$ are fixed, $q \to \infty$ and $q \to 0$ correspond
to $\tau_c \to 0$ or $\tau_c \to \infty$, that is, very quick or very slow variation
of $\bu$ in time.
When discussing the limits $q \to \infty$ or $q \to 0$ in the following we always consider
$\lambda_c$ as finite and $\tau_c$ as non--zero.

The second--order correlation approximation applies as soon as the neglect of $\bG$
in (\ref{eq107}) is justified.
A sufficient condition for that is $|\bbb| \ll |\bmB|$.
If $q \lesssim 1$ this is equivalent to $Rm \ll 1$, if $q \gtrsim 1$ to $St \ll 1$.
We may sum up this by saying that a sufficient condition reads
\begin{equation}
\mbox{min} \, (Rm, St) \ll 1
\label{eq222}
\end{equation}
(see also, e.g., Krause and R\"adler 1971, 1980 or R\"adler 2000).
Incidentally, determinations of $a_{ij}$ and $b_{ijk}$ have been
done within the framework of numerical simulations of
magnetoconvection or dynamos without using the second--order
correlation approximation. It turned out that, e.g., for $q
\lesssim 1$ there is a rather good agreement of the results for
these quantities with those obtained in the second--order
correlation approximation not only for $Rm \ll 1$ but for values
of $Rm$ up to about 10
\citep{schrinneretal05,schrinneretal06,schrinneretal07}.

\subsubsection{Limiting cases concerning $q$}

Let us specify the above results for $a_{ip}$ and $b_{ipq}$ to the limiting cases with respect to $q$.
As shown in appendix~\ref{app1}, in the limit $q \to \infty$ the relations (\ref{eq213})
for $a_{ip}$ and $b_{ipq}$ turn into
\begin{eqnarray}
a_{ip} &=& (\epsilon_{ijn} \delta_{lp} - \epsilon_{ijp} \delta_{ln}) \, \int_0^\infty \langle u_j (\bx, t)
    \frac{\p u_n (\bx, t - \tau)}{\p x_l} \rangle \, \dd \tau
\label{eq227}\\
&& = \epsilon_{ijn} \Big( \int_0^\infty \langle u_j (\bx, t) \frac{\p u_n (\bx, t - \tau)}{\p x_p} \rangle \, \dd \tau
     - \delta_{np} \int_0^\infty \langle u_j (\bx, t) \big(\bnab \cdot \bu (\bx, t - \tau)\big) \rangle \, \dd \tau \Big)
\nonumber\\
b_{ipq} &=& - \epsilon_{ijp} \int_0^\infty
     \langle u_j (\bx, t) \, u_q (\bx, t - \tau) \rangle \, \dd \tau  \, .
\nonumber
\end{eqnarray}
A contribution to $b_{ipq}$ containing $\delta_{pq}$,
which is because of $\bnab \cdot \bmB = 0$ meaningless for $\bscE$, has been omitted.

In the limit $q \to 0$ we obtain
\begin{eqnarray}
a_{ip} &=& \frac{\epsilon_{ijn} \delta_{lp} - \epsilon_{ijp} \delta_{ln}}{4 \pi \eta} \,
    \int_\infty \langle u_j (\bx, t) \, u_n (\bx + \bxi, t) \rangle \, \xi_l \frac{\dd^3 \xi}{\xi^3}
\nonumber\\
&&  = \frac{\epsilon_{ijn}}{4 \pi \eta} \, \Big( \int_\infty \langle u_j (\bx, t)
    \frac{\p u_n (\bx + \bxi, t)}{\p x_p} \rangle \, \frac{\dd^3 \xi}{\xi}
    - \delta_{np} \, \int_\infty \langle u_j (\bx, t) \big(\bnab \cdot \bu (\bx + \bxi, t)\big) \rangle \,
    \frac{\dd^2 \xi}{\xi} \, \Big)
\label{eq231}\\
b_{ipq} &=& \frac{1}{4 \pi \eta} \, \int_\infty
     (\epsilon_{ijn} \xi_p - \epsilon_{ijp} \xi_n) \, \xi_q
     \langle u_j (\bx, t) \, u_n (\bx + \bxi, t) \rangle \, \frac{\dd^3 \xi}{\xi^3}   \, .
\nonumber
\end{eqnarray}

We note that these results for the limits $q \to \infty$ and $q \to 0$
can also be derived from equation (\ref{eq200}) with $\eta \nabla^2 \bbb$ or $\p_t \bbb$,
respectively, cancelled
\citep{krauseetal71b,krauseetal80}.
Interestingly enough, for the limit $q \to \infty$ then,
without introducing (\ref{eq209}) or similar assumptions,
no contributions to $\bscE$ with higher than first--order spatial derivatives
of $\bmB$ occur.
This is in agreement with the fact that, when considering first the general result for $\bscE$
with higher spatial derivatives (R\"adler 1968a,b)
and proceeding then to the limit $q \to \infty$
all the coefficients of the higher than first--order spatial derivatives vanish.
This finding might give some justification for considering no contributions
to $\bscE$ with higher than first--order spatial derivatives
in astrophysical applications,
in which the limit $q \to \infty$ is a good approximation.
There is, however, no general justification to ignore contributions to $\bscE$
with time derivatives of $\bmB$.
By contrast, in the case $q \to 0$ contributions to $\bscE$ with higher spatial derivatives of $\bmB$
are well possible but no contributions with time derivatives of $\bmB$.

\subsubsection{Homogeneous isotropic turbulence}
\label{subsubsec313}

It is instructive to consider the special case of a homogeneous isotropic turbulence,
for which $\bscE$ takes the form (\ref{eq113}), that is $\bscE = \alpha \, \bmB - \beta \, \bnab \x \bmB$.
For $\alpha$ and $\beta$, which are independent of position, we find then from (\ref{eq213})
\begin{eqnarray}
\alpha &=& - \frac{1}{3} \int_\infty \int_0^\infty \frac{\p G^{(\eta)} (\xi, \tau)}{\p \xi} \,
     \langle \bu (\bx, t) \x \bu (\bx + \bxi, t - \tau) \rangle \cdot (\bxi/\xi) \, \dd^3 \xi \, \dd \tau
\nonumber\\
&=& - \frac{1}{3} \int_\infty \int_0^\infty G^{(\eta)} (\xi, \tau) \,
     \langle \bu (\bx, t) \cdot \big(\bnab \x \bu (\bx + \bxi, t - \tau)\big) \rangle \, \dd^3 \xi \, \dd \tau
\label{eq237}\\
\beta &=& - \frac{1}{9} \, \int_\infty \int_{0}^\infty \frac{\p G^{(\eta)} (\xi, \tau)}{\p \xi}
     \langle \bu (\bx, t) \cdot \bu (\bx + \bxi, t - \tau) \rangle \, \xi \, \dd^3 \xi \, \dd \tau \, .
\nonumber
\end{eqnarray}
Due to the assumed isotropy of the turbulence we may replace
$\langle \bu (\bx, t) \cdot \bu (\bx + \bxi, t - \tau) \rangle$
by $3 \langle u_\xi (\bx, t) \, u_\xi (\bx + \bxi, t - \tau) \rangle$,
where $u_\xi = (\bu \cdot \bxi) / \xi$.
Incidentally, since $G^{(\eta)}$ and all averages $\langle \ldots \rangle$ depend on $\bxi$ via $\xi$ only
we may replace the integrals $\int_\infty \ldots \dd^3 \xi$ by $4 \pi \, \int_0^\infty \ldots \xi^2 \dd \xi$.

In the limit $q \to \infty$ we have
\begin{eqnarray}
\alpha &=& - \, \frac{1}{3} \int_0^\infty
    \langle \bu (\bx, t) \cdot \big(\bnab \x \bu (\bx, t - \tau)\big) \rangle \, \dd \tau
\nonumber\\
\beta &=& \frac{1}{3} \int_0^\infty
    \langle \bu (\bx, t) \cdot \bu (\bx, t - \tau) \rangle  \, \dd \tau \, .
\label{eq241}
\end{eqnarray}
This result is often written in the form
\begin{equation}
\alpha = - \, \frac{1}{3}
    \langle \bu \cdot (\bnab \x \bu) \rangle \, \tau^{(\alpha)} \, , \quad
    \beta = \frac{1}{3}
    \langle \bu ^2 \rangle \, \tau^{(\beta)} \, .
\label{eq243}
\end{equation}
Here $\tau^{(\alpha)}$ and $\tau^{(\beta)}$ are just defined by equating
the right--hand sides of (\ref{eq241})
to those of (\ref{eq243}), respectively.
This is, of course, only possible under the reasonable assumption that
$\langle \bu (\bx, t) \cdot \big(\bnab \x \bu (\bx, t - \tau)\big) \rangle$ at $\tau = 0$ is unequal to zero.
It seems plausible to assume that both $\tau^{(\alpha)}$ and $\tau^{(\beta)}$
are of the order of the correlation time $\tau_c$ of the turbulent velocity field.
There is, however, hardly a general reason for their exact equality.

Remarkably, as explained in appendix~\ref{app2}, in the case of an
incompressible fluid and statistically steady turbulence the
right--hand sides of (\ref{eq241}) can also be expressed by integrals
containing $\langle \bu (\bx, t) \cdot \big(\bnab \x \bu (\bx + \bxi, t)\big) \rangle$
or $\langle \bu (\bx, t) \cdot \bu (\bx + \bxi, t) \rangle$, taken over all $\xi$.

In the limit $q \to 0$ we obtain
\begin{eqnarray}
\alpha &=& \frac{1}{12 \pi \eta} \int_\infty
    \big(\langle \bu (\bx, t) \x \bu (\bx + \bxi, t) \rangle \cdot \bxi \big) \,
    \frac{\dd^3 \xi}{\xi^3}
    = - \, \frac{1}{12 \pi \eta} \int_\infty
    \langle \bu (\bx, t) \cdot \big(\bnab \x \bu (\bx + \bxi, t)\big) \rangle \,
    \frac{\dd^3 \xi}{\xi}
\nonumber\\
\beta &=& \frac{1}{36 \pi \eta} \int_\infty
    \langle \bu (\bx, t) \cdot \bu (\bx + \bxi, t) \rangle \,
    \frac{\dd^3 \xi}{\xi} \, .
\label{eq247}
\end{eqnarray}
Again the integrals $\int_\infty \ldots\, \dd^3 \xi$ may be replaced by
$4 \pi \, \int_0^\infty \ldots\, \xi^2 \dd \xi$.

Incidentally, the results (\ref{eq247}) take a simple form
if we represent $\bu$ by $\bu = \bnab \x \bpsi - \bnab \phi$
with the vector potential $\bpsi$ and a scalar potential $\phi$.
They read then
\begin{equation}
\alpha = - \, \frac{1}{3 \eta} \, \langle \bu \cdot \bpsi \rangle
    = - \, \frac{1}{3 \eta} \, \langle \bpsi \cdot (\bnab \x \bpsi) \rangle \, , \quad
\beta = \frac{1}{3 \eta} \, (\langle \bpsi^2 \rangle - \langle \phi^2 \rangle )
\label{eq249}
\end{equation}
(see, e.g., Krause and R\"adler 1971, 1980,
but note that there is a sign error in Krause and R\"adler 1971).

\subsubsection{Comments}
\label{subsubsec314}

In the above calculation of the electromotive force $\bscE$ no other restrictions
concerning the fluctuating fluid velocity $\bu$ have been used than some ``smallness"
which ensures the applicability of the second--order correlation approximation.

We note that in this approximation if, as so far done, a purely hydrodynamic turbulence
is considered  and any influence of the initial $\bbb$ is ignored,
$\bscE$ does not contain a part $\bscE^{(0)}$.

We may of course specify the velocity $\bu$ or its correlation tensor $Q_{ij}$
such that they correspond to a turbulence on a rotating body,
that is, subject to a Coriolis force.
The dependence of $\bu$ and $Q_{ij}$ on the angular velocity $\bOmega$
responsible for the Coriolis force has to be calculated
on the basis of the momentum balance (\ref{eq141}).
Likewise we may consider a turbulence under the influence of a given mean magnetic field, that is,
under the action of the Lorentz force, and specify $\bu$, again on the basis of the momentum balance,
to be a function of $\bmB$ and its derivatives.
In that sense the above derivations may serve also as a starting point for investigations
of the quenching of the mean--field induction effects by the mean magnetic field.

As for higher than second--order correlation approximations we refer to section~\ref{new1}.

\subsection{Magnetohydrodynamic background turbulence}
\label{subsec32}

\subsubsection{Second--order correlation approximation}
\label{subsubsec321}

Let us relax the assumption that $\bbb$ decays to zero if $\bmB$ vanishes, that is,
admit a magnetohydrodynamic background turbulence.
We assume the fluid to be incompressible.
In addition to the induction equation (\ref{eq107}) we use now
from the very beginning also the momentum balance in the form (\ref{eq141}).
For the sake of simplicity we restrict our attention to a non--rotating fluid,
that is $\bOmega = \bzo$, and exclude again any mean motion, $\bmU = \bzo$.
Then $\bu$ and $\bbb$ are governed by
\begin{eqnarray}
\p_t \bu  &=&  \nu \nabla^2 \bu
    + \frac{1}{\mu \varrho} \big( (\bmB \cdot \bnab) \bbb + (\bbb \cdot \bnab) \bmB + \bnab M \big)
    + \tilde{\bff} + \tilde{\bT} \, , \quad
    \bnab \cdot \bu = 0
\nonumber\\
\p_t \bbb &=& \eta \nabla^2 \bbb + (\bmB \cdot \bnab) \bu - (\bu \cdot \bnab) \bmB  + \bG \, , \quad
    \bnab \cdot \bbb = 0
\label{eq261}
\end{eqnarray}
with $M$, $\tilde{\bff}$ and $\tilde{\bT}$ as defined in (\ref{eq143}).

We expand $\bu$ and $\bbb$ with respect to $\bmB$,
\begin{equation}
\bu = \bu^{(0)} + \bu^{(1)} +  \ldots \, , \quad
    \bbb = \bbb^{(0)} + \bbb^{(1)} +  \ldots \, ,
\label{equ263}
\end{equation}
where $\bu^{(0)}$ and $\bbb^{(0)}$ are independent of $\bmB$,
further $\bu^{(1)}$ and $\bbb^{(1)}$ of first order in $\bmB$,
and $\ldots$ stands for terms of higher order in $\bmB$
(see also Blackman and Field 1999 or Field {\it et al.} 1999).
Then we have
\begin{eqnarray}
\bscE &=& \bscE^{(0)} + \bscE^{(1)} + \ldots \, , \quad
     \bscE^{(1)} = \bscE^{(01)} + \bscE^{(10)} \, , \quad \ldots
\nonumber\\
\bscE^{(0)} &=& \langle \bu^{(0)} \x \bbb^{(0)} \rangle \, , \quad
     \bscE^{(01)} = \langle \bu^{(0)} \x \bbb^{(1)} \rangle \, , \quad
     \bscE^{(10)} = \langle \bu^{(1)} \x \bbb^{(0)} \rangle \, , \quad \ldots \, .
\label{eq265}
\end{eqnarray}

We split now the equations (\ref{eq261}) in the usual way into equations for $\bu^{(0)}$ and $\bbb^{(0)}$,
which do not contain $\bu^{(1)}$ and $\bbb^{(1)}$, and equations for $\bu^{(1)}$ and $\bbb^{(1)}$,
which contain of course also $\bu^{(0)}$ and $\bbb^{(0)}$.
We consider those for $\bu^{(0)}$ and $\bbb^{(0)}$ as being solved,
that is, $\bu^{(0)}$ and $\bbb^{(0)}$ as given.
When deriving the equations for $\bu^{(1)}$ and $\bbb^{(1)}$ we assume that $\bff$ is independent of $\bmB$.
We further use the second--order approximation, in this context understood
as the neglect of $\bG^{(1)}$ and ${\tilde{\bT}}^{(1)}$ in these equations.
Here $\bG^{(1)}$ is defined by
$\bG^{(1)} = \bnab \x \big((\bu^{(0)} \x  \bbb^{(1)})' + (\bu^{(1)} \x  \bbb^{(0)})'\big)$
and ${\tilde{\bT}}^{(1)}$ is a quantity derived in the sense of (\ref{eq129}) and (\ref{eq143})
from the analogously defined $\bT^{(1)}$.
In this way we arrive at
\begin{eqnarray}
\p_t \bu^{(1)} - \nu \nabla^2 \bu^{(1)} &=& \frac{1}{\mu \varrho}
     \big( (\bmB \cdot \bnab) \bbb^{(0)} + (\bbb^{(0)} \cdot \bnab) \bmB + \bnab M^{(1)} \big) \, , \quad
     \bnab \cdot \bu^{(1)} = 0
\nonumber\\
\p_t \bbb^{(1)} - \eta \nabla^2 \bbb^{(1)} &=& (\bmB \cdot \bnab) \bu^{(0)} - (\bu^{(0)} \cdot \bnab) \bmB \, , \quad
     \bnab \cdot \bbb^{(1)} = 0 \, .
\label{eq267}
\end{eqnarray}
$M^{(1)}$ is defined as $M$ in (\ref{eq143}) but with $\bbb^{(0)}$ instead of $\bbb$.

We recall here the derivations of section~\ref{subsubsec311}.
Starting from equation (\ref{eq200})
we have there calculated ${\cE}_i$ in the form (\ref{eq207}).
With the second equation (\ref{eq267}) we find on the same way
\begin{equation}
{\cE}^{(01)}_i (\bx, t) =  \big( \epsilon_{ijn} \delta_{lp} - \epsilon_{ijp} \delta_{ln} \big) \,
     \int_\infty \int_{0}^\infty \frac{\p G^{(\eta)} (\xi, \tau)}{\p \xi}
     \frac{\xi_l}{\xi} \, Q^{(0)}_{jn} (\bx, t; - \bxi, - \tau) \, \mB_p (\bx - \bxi, t - \tau) \,
     \dd^3 \xi \, \dd \tau \, .
\label{eq269}
\end{equation}
$Q_{ij}^{(0)}$ is in analogy to $Q_{ij}$ defined by
\begin{equation}
Q^{(0)}_{ij} (\bx, t; \bxi, \tau) = \langle u^{(0)}_i (\bx, t) \, u^{(0)}_j (\bx + \bxi, t + \tau) \rangle \, .
\label{eq271}
\end{equation}
Because of $\bnab \cdot \bu^{(0)} = 0$ we have $\p Q^{(0)}_{ij}/ \p x_i - \p Q^{(0)}_{ij}/ \p \xi_i = 0$
and $\p Q^{(0)}_{ij}/ \p \xi_j = 0$.

The first equation (\ref{eq267}) leads to
\begin{eqnarray}
{\cE}^{(10)}_i (\bx, t) &=&  - \, \frac{\epsilon_{ijn} \, \delta_{lp} + \epsilon_{ijp} \, \delta_{ln}}{\mu \varrho } \,
     \int_\infty \int_{0}^\infty \frac{\p G^{(\nu)} (\xi, \tau)}{\p \xi} \frac{\xi_l}{\xi} \,
     R^{(0)}_{jn} (\bx, t; - \bxi, - \tau) \, \mB_p (\bx - \bxi, t - \tau) \, \dd^3\xi \dd \tau
\nonumber\\
&& \qquad \qquad \qquad
     - \, \frac{2 \, \epsilon_{ijl}}{\mu \varrho}
     \int_\infty \int_{0}^\infty \frac{\p G^{(\nu)} (\xi, \tau)}{\p \xi} \frac{\xi_l}{\xi} \,
     N^{(0)}_j (\bx, t; - \bxi, - \tau) \, \dd^3 \xi \, \dd \tau \, .
\label{eq273}
\end{eqnarray}
$G^{(\nu)} (\xi, \tau)$ is defined like $G^{(\eta)} (\xi, \tau)$ in (\ref{eq203}),
but with $\eta$ replaced by $\nu$.
Further $R_{ij}^{(0)}$ is in analogy to $Q_{ij}^{(0)}$ defined by
\begin{equation}
R_{ij}^{(0)} (\bx, t; \bxi, \tau) = \langle b_i^{(0)} (\bx, t) \, b_j^{(0)} (\bx + \bxi, t + \tau) \rangle
\label{eq275}
\end{equation}
and $N^{(0)}_j$ by
\begin{equation}
N^{(0)}_j (\bx, t; \bxi, \tau) = \frac{1}{4 \pi} \int_\infty
     \frac{\p R^{(0)}_{jn} (\bx, t; \bxi', \tau)}{\p {\xi'}_p} \,
     \frac{\p \mB_p (\bx + \bxi', t + \tau)}{\p {\xi'}_n} \, \frac{\dd^3 \xi'}{| \bxi - \bxi' |} \, .
\label{eq276}
\end{equation}
Because of $\bnab \cdot \bbb^{(0)} = 0$ we have $\p R^{(0)}_{ij}/ \p x_i - \p R^{(0)}_{ij}/ \p \xi_i = 0$
and $\p R^{(0)}_{ij}/ \p \xi_j = 0$.

We restrict our attention now to the limit of small $\bmB$.
As in section~\ref{subsubsec311} we assume in addition sufficiently weak variations of $\bmB$ in $\bx$ and $t$
and introduce the truncated Taylor expansion (\ref{eq209}) into the integrands
of (\ref{eq269}), (\ref{eq273}) and (\ref{eq276}).
Considering (\ref{eq265}) and ignoring contributions to $\bscE$ of higher than first order in $\bmB$
we find then again (\ref{eq111}),
that is ${\cE}_i = {\cE}^{(0)}_i + a_{ij} \, \mB_j + b_{ijk} \, \p \mB_j / \p x_k$.
Here we have $\bscE^{(0)} = \langle \bu^{(0)} \x \bbb^{(0)} \rangle$,
and this may in general well be unequal to zero.
We put
\begin{equation}
a_{ij} = a^{(u)}_{ij} + a^{(b)}_{ij} \, , \quad
    b_{ijk} = b^{(u)}_{ijk} + b^{(b)}_{ijk} \, ,
\label{eq279}
\end{equation}
with $a^{(u)}_{ij}$ and $b^{(u)}_{ijk}$ determined by $\bu^{(0)}$,
and $a^{(b)}_{ij}$ and $b^{(b)}_{ijk}$ by $\bbb^{(0)}$.
Then we have
\begin{eqnarray}
a^{(u)}_{ip} &=&  \epsilon_{ijk}
     \int_\infty \int_{0}^\infty G^{(\eta)} (\xi, \tau)\,
     \frac{\p Q_{jk}^{(0)} (\bx, t; \bxi, - \tau)}{\p \xi_p} \, \dd^3 \xi \, \dd \tau
\nonumber\\
&=&  - \, \epsilon_{ijk}
     \int_\infty \int_{0}^\infty \frac{\p G^{(\eta)} (\xi, \tau)}{\p \xi}
     \frac{\xi_p}{\xi} Q^{(0)}_{jk} (\bx, t; \bxi, - \tau) \, \dd^3 \xi \, \dd \tau
\nonumber\\
a^{(b)}_{ip} &=& - \frac{\epsilon_{ijk}}{\mu \varrho}
     \int_\infty \int_{0}^\infty G^{(\nu)} (\xi, \tau)\,
     \frac{\p R_{jk}^{(0)} (\bx, t; \bxi, - \tau)}{\p \xi_p} \, \dd^3 \xi \, \dd \tau
\nonumber\\
&=& \frac{\epsilon_{ijk}}{\mu \varrho}
     \int_\infty \int_{0}^\infty \frac{\p G^{(\nu)} (\xi, \tau)}{\p \xi}
     \frac{\xi_p}{\xi} R^{(0)}_{jk} (\bx, t; \bxi, - \tau) \, \dd^3 \xi \, \dd \tau
\nonumber\\
b^{(u)}_{ipq} &=& - \, \epsilon_{ijn} \, \int_\infty \int_{0}^\infty \frac{\p G^{(\eta)} (\xi, \tau)}{\p \xi}
     \frac{\xi_p \xi_q}{\xi} Q^{(0)}_{jn} (\bx, t; \bxi, - \tau) \, \dd^3 \xi \, \dd \tau
\label{eq281}\\
&&  \qquad - \, \epsilon_{ijp} \int_\infty \int_{0}^\infty G^{(\eta)} (\xi, \tau)
     Q^{(0)}_{jq} (\bx, t; \bxi, - \tau) \, \dd^3 \xi \, \dd \tau
\nonumber\\
b^{(b)}_{ipq} &=& \frac{\epsilon_{ijn}}{\mu \varrho} \,
     \int_\infty \int_{0}^\infty \frac{\p G^{(\nu)} (\xi, \tau)}{\p \xi}
     \frac{\xi_p \xi_q}{\xi} R^{(0)}_{jn} (\bx, t; \bxi, - \tau) \, \dd^3 \xi \, \dd \tau
\nonumber\\
&&  \qquad - \, \frac{1}{\mu \varrho} \int_\infty \int_{0}^\infty G^{(\nu)} (\xi, \tau)
     \big( \epsilon_{ijp} - 2 \, \epsilon_{ijl} \, \frac{\p^2}{\p \xi_l \p \xi_p} \Delta_\xi^{-1} \big)
     R^{(0)}_{jq} (\bx, t; \bxi, - \tau) \, \dd^3 \xi \, \dd \tau \, ,
\nonumber
\end{eqnarray}
where
\begin{equation}
\Delta_\xi^{-1} F (\bx, t; \bxi, \tau) = - \frac{1}{4 \pi}
\int_\infty
     \frac{F (\bx, t; \bxi', \tau)}{|\bxi - \bxi'|} \dd^3 \xi' \, .
\label{eq283}
\end{equation}
By the way, the above relations for $a^{(u)}_{ip}$ and $b^{(u)}_{ipq}$
can also be concluded from (\ref{eq213}) by replacing $a_{ip}$, $b_{ipq}$ and $Q_{jn}$
with $a^{(u)}_{ip}$, $b^{(u)}_{ipq}$ and $Q^{(0)}_{jn}$, respectively,
and considering $\p Q^{(0)}_{ij}/ \p \xi_j = 0$.

\subsubsection{Range of validity}
\label{subsubsec322}

In addition to the dimensionless parameters $Rm$, $St$ and $q$ introduced with (\ref{eq220}) and (\ref{eq221})
we define the hydrodynamic Reynolds number $Re$ and a parameter $p$ by
\begin{equation}
Re = u_c \lambda_c / \nu \, , \quad p = \lambda^2_c / \nu \tau_c \, .
\label{eq287}
\end{equation}
We might call the cases $p \to \infty$ and $p \to 0$ ``low--viscosity limit" and ``high--viscosity limit".
However, the remarks made on the notations ``high-conductivity limit" and ``low--conductivity limit"
made under (\ref{eq221}) apply analogously.
As in the case of $q$, when discussing the limits $p \to \infty$ or $p \to 0$
in the following we always consider $\lambda_c$ as finite and $\tau_c$ as non--zero.

Modifying the reasoning which lead us to the condition (\ref{eq222}) properly
we find that a sufficient condition for the validity of our above results
for magnetohydrodynamic background turbulence is given by
\begin{equation}
\mbox{min} (Rm, St) \ll 1 \, , \quad \mbox{min} (Re, St) \ll 1  \, .
\label{eq288}
\end{equation}

\subsubsection{A limiting case}

We note that $a^{(u)}_{ij}$ and $b^{(u)}_{ijk}$ depend only on $q$ and not on $p$,
and $a^{(b)}_{ij}$ and $b^{(b)}_{ijk}$ only on $p$ and not on $q$.
The relations for $a^{(u)}_{ij}$ and $b^{(u)}_{ijk}$ in the limits $q \to \infty$ and $q \to 0$
agree with those for $a_{ij}$ and $b_{ijk}$ given with (\ref{eq227}) and (\ref{eq231}),
specified to $\bnab \cdot \bu = 0$ and with $\bu$ is replaced by $\bu^{(0)}$.
For $a^{(b)}_{ij}$ and $b^{(b)}_{ijk}$ we find on the way described in appendix~\ref{app1}
in the limit $p \to \infty$
\begin{eqnarray}
a^{(b)}_{ip} &=& - \, \frac{\epsilon_{ijk}}{\mu \varrho} \int_0^\infty
     \langle b_j^{(0)} (\bx, t) \, \frac{\p b_k^{(0)} (\bx, t - \tau)}{\p x_p} \, \rangle \, \dd \tau
\nonumber\\
b^{(b)}_{ipq} &=& - \frac{\epsilon_{ijp}}{\mu \varrho}
     \int_0^\infty \langle b_j^{(0)} (\bx, t) \, b_q^{(0)} (\bx, t - \tau) \, \rangle \, \dd \tau
\label{eq289}\\
&& - \, \frac{\epsilon_{ijl}}{2 \pi \mu \varrho} \, \int_\infty \int_0^\infty
     \frac{\p^2}{\p \xi_l \p \xi_p} \Big( \frac{1}{\xi}\Big) \,
     \langle b_j^{(0)} (\bx, t) \, b_q^{(0)} (\bx + \bxi, t - \tau) \, \rangle \, \dd^3 \xi \, \dd \tau \, .
\nonumber
\end{eqnarray}
A contribution to $b^{(b)}_{ipq}$ containing $\delta_{pq}$ has been omitted.
In the limit $p \to 0$ we have
\begin{eqnarray}
a^{(b)}_{ip} &=& - \, \frac{\epsilon_{ijk}}{4 \pi \mu \varrho \, \nu} \int_\infty
     \langle b_j^{(0)} (\bx, t) \, b_k^{(0)} (\bx + \bxi, t) \rangle \, \xi_p \frac{\dd^3 \xi}{\xi^3}
     = - \frac{\epsilon_{ijk}}{4 \pi \mu \varrho \, \nu} \int_\infty
     \langle b_j^{(0)} (\bx, t) \, \frac{\p b_k^{(0)} (\bx + \bxi, t)}{\p x_p} \rangle \, \frac{\dd^3 \xi}{\xi}
\nonumber\\
b^{(b)}_{ipq} &=&  - \, \frac{\epsilon_{ijm}}{4 \pi \mu \varrho \, \nu} \int _\infty
     \big(\delta_{mp} \delta_{nq} + \delta_{mn} \frac{\xi_p \xi_q}{\xi^2} \big) \,
     \langle b_j^{(0)} (\bx, t) \, b_n^{(0)} (\bx + \bxi, t) \rangle \, \frac{\dd^3 \xi}{\xi}
\label{eq291}\\
&& - \, \frac{\epsilon_{ijl}}{8 \pi^2 \mu \varrho \, \nu} \int_\infty \int_\infty
     \frac{\p^2}{\p \xi'_l \p \xi'_p} \Big( \frac{1}{|\xi - \xi'|} \Big) \,
     \langle b_j^{(0)} (\bx, t) \, b_q^{(0)} (\bx + \bxi', t) \rangle \,
     \dd^3 \xi' \, \frac{\dd^3 \xi}{\xi} \, .
\nonumber
\end{eqnarray}

\subsubsection{Homogeneous isotropic turbulence}
\label{subsubsec323}

Let us now again consider the special case of homogeneous isotropic magnetohydrodynamic turbulence.
Then $\bscE$ takes again the form (\ref{eq113}), that is $\bscE = \alpha \, \bmB - \beta \, \bnab \x \bmB$.
On the basis of (\ref{eq279}) and (\ref{eq281}) we find
\begin{equation}
\alpha = \alpha^{(u)} - \alpha^{(b)} \, , \quad \beta = \beta^{(u)}
\label{eq297}
\end{equation}
with
\begin{eqnarray}
\alpha^{(u)} &=& - \frac{1}{3} \int_\infty \int_0^\infty \frac{\p G^{(\eta)} (\xi, \tau)}{\p \xi} \,
     \langle \bu^{(0)}(\bx, t) \x \bu^{(0)}(\bx + \bxi, t - \tau) \rangle \cdot (\bxi/\xi) \, \dd^3 \xi \, \dd \tau
\nonumber\\
&=& - \frac{1}{3} \int_\infty \int_0^\infty G^{(\eta)} (\xi, \tau)
    \langle \bu^{(0)} (\bx, t) \cdot \big(\bnab \x \bu^{(0)} (\bx + \bxi, t - \tau)\big) \rangle \dd^3 \xi \dd \tau
\nonumber\\
\alpha^{(b)} &=& - \frac{1}{3 \mu \varrho} \, \int_\infty \int_{0}^\infty \frac{\p G^{(\nu)} (\xi, \tau)}{\p \xi}
       \langle \bbb^{(0)}(\bx, t) \x \bbb^{(0)}(\bx + \bxi, t - \tau) \rangle \cdot (\bxi/\xi) \, \dd^3 \xi \, \dd \tau
\label{eq299}\\
&=& - \frac{1}{3 \mu \varrho} \int_\infty \int_0^\infty G^{(\nu)} (\xi, \tau)
    \langle \bbb^{(0)} (\bx, t) \cdot \big(\bnab \x \bbb^{(0)} (\bx + \bxi, t - \tau)\big) \rangle \dd^3 \xi \dd \tau
\nonumber\\
\beta^{(u)} &=& \frac{1}{3} \int_\infty \int_0^\infty G^{(\eta)} (\xi, \tau)
    \langle \bu^{(0)} (\bx, t) \cdot \bu^{(0)} (\bx + \bxi, t - \tau) \rangle \dd^3 \xi \dd \tau \, .
\nonumber
\end{eqnarray}
By the reason explained in the context of (\ref{eq237}) the integrals $\int_\infty \ldots \dd^3 \xi$
in (\ref{eq299}) may be replaced by $4 \pi \int_0^\infty \ldots \xi^2 \dd \xi$.
Interestingly enough there is no ``magnetic part" of $\beta$.
This agrees with a result by \citet{vainshteinetal83} obtained within an early version
of the $\tau$--approach applying to the limit $q \to \infty$.

In the case $q, p \to \infty$ we have
\begin{eqnarray}
\alpha^{(u)} &=& - \frac{1}{3} \int_0^\infty
    \langle \bu^{(0)} (\bx, t) \cdot \big(\bnab \x \bu^{(0)} (\bx, t - \tau)\big) \rangle \dd \tau
\nonumber\\
\alpha^{(b)} &=& - \frac{1}{3 \mu \varrho} \int_0^\infty
    \langle \bbb^{(0)} (\bx, t) \cdot \big(\bnab \x \bbb^{(0)} (\bx, t - \tau)\big) \rangle \dd \tau
\label{eq301}\\
\beta^{(u)} &=& \frac{1}{3} \int_0^\infty
    \langle \bu^{(0)} (\bx, t) \cdot \bu^{(0)} (\bx, t - \tau) \rangle \dd \tau \, .
\nonumber
\end{eqnarray}
In the sense discussed in the context of (\ref{eq243}) we write this also in the form
\begin{eqnarray}
\alpha^{(u)} &=& - \frac{1}{3} \langle \bu^{(0)} \cdot (\bnab \x \bu^{(0)}) \rangle \, \tau^{(\alpha \, u)} \, , \quad
   \alpha^{(b)} = - \frac{1}{3 \mu \varrho} \langle \bbb^{(0)} \cdot (\bnab \x \bbb^{(0)}) \rangle \, \tau^{(\alpha \, b)}
\nonumber\\
\beta^{(u)} &=& \frac{1}{3} \langle {\bu^{(0)}}^2 \rangle \, \tau^{(\beta \, u)}
\label{eq303}
\end{eqnarray}
with times $\tau^{(\alpha \, u)}$, $\tau^{(\alpha \, b)}$ and $\tau^{(\beta \, u)}$
defined by equating the corresponding right--hand sides of (\ref{eq301}) and (\ref{eq303}).

\subsubsection{Comments}
\label{subsubsec324}

In the results for purely hydrodynamic background turbulence presented in section~\ref{subsec31}
the velocity $\bu$ may well depend on $\bmB$.
If we ignore this dependency $\bu$ agrees with $\bu^{(0)}$ introduced here.
In this case the results of section~\ref{subsec31} coincide with those obtained here
if specified to $\bbb^{(0)} = \bzo$.
We stress that in the case of a purely hydrodynamic background turbulence,
even if we assume that $\bu$ is subject to an influence of the Lorentz force and so depends on $\bmB$,
e.g., the relations (\ref{eq213}), (\ref{eq227}) and (\ref{eq231}) for $a_{ip}$ remain their validity
and have not to be modified by adding terms which explicitly depend on $\bbb$.
Only if we replace there $\bu$ by $\bu^{(0)}$ an additional term containing $\bbb$
might occur.

The relation for $\alpha$ defined by (\ref{eq297}) and (\ref{eq303}) is usually taken
to be supported by the analysis of \citet{pouquetetal76},
who employ a much more sophisticated closure than the $\tau$--approaches,
that is, the ``eddy damped quasi-normal Markovian approximation".
We note however that the justification given there (subsection ``Phenomenology
of the M-helicity effect" of their section 3) is based on the assumption of an
(at least initially) purely magnetic background turbulence, see their equation (3.19).
It has to be considered as a consequence of a given electromotive force
rather than a kinetic driving force.
Under these conditions, too, it is not surprising to see a term depending
on the current helicity $\langle \bbb \cdot (\bnab \x \bbb) \rangle$
in the expression for $\alpha$, just as in our above derivation
leading to the results (\ref{eq297}) and (\ref{eq303}).

\section{The $\tau$--approaches (approaches (ii))
to the mean electromotive force $\bscE$}
\label{secii}

As mentioned above, several analytical approaches to the mean electromotive force $\bscE$
going beyond the second--order correlation approximation have been proposed
which make use of some modification of the $\tau$--approximation of the hydrodynamic turbulence theory
\citep{orszag70}.
In contrast to approach (i), in these approaches, labelled by (ii) in the following,
both the induction equation (\ref{eq107})
and the momentum balance in the form (\ref{eq141}) with the forcing term $\tilde{\bff}$
and the non--linear terms $\bG$ and $\tilde{\bT}$ are used from the very beginning.
In this way also $\bnab \cdot \bu = 0$ is introduced.
Equations (\ref{eq107}) and (\ref{eq141}) are however not really solved.
Instead, relations for $\bscE$ are derived which contain, of course,
contributions resulting from the mentioned quantities, that is, $\tilde{\bff}$, $\bG$ and $\tilde{\bT}$.
For these contributions then some kind of $\tau$--approximation is introduced.
In this section we give a critical review of the approaches of type (ii).
In particular we check to which extent their results satisfy the elementary requirement
that they agree with results of approach (i) in the range of its validity.

For the sake of simplicity we exclude again any mean motion of the fluid, that is, $\bmU = \bzo$.
We further restrict our attention to the case of a non--rotating fluid, that is, $\bOmega = \bzo$,
and give only short comments on the case with $\bOmega \not= \bzo$.

\subsection{The simple $\tau$--approach
(approach (iia))}
\label{subseciia}

Let us first describe a simple approach to the mean electromotive force $\bscE$,
which we call approach (iia) in the following.
It defines a frame which allows us to explain and to discuss the approaches used
by \citet{vainshteinetal83}, \citet{blackmanetal02b} and \citet{brandenburgetal05b}.

\subsubsection{Outline of approach (iia)}
\label{subsubseciia1}

We calculate first the time derivative of $\bscE$ on the basis of
\begin{equation}
\p_t \bscE = \langle \p_t \bu \x \bbb + \bu \x \p_t \bbb \rangle \, .
\label{eq311}
\end{equation}
Using equations (\ref{eq107}), further (\ref{eq141}) with $\tilde{\bR} = \tilde{\bU} = \bzo$,
and considering $\p \mB_p / \p x_q$ as constant so that (\ref{eq145}) applies, we obtain
\begin{equation}
\p_t \bscE = \bI + \bY + \bZ \, ,
\label{eq313}
\end{equation}
where
\begin{eqnarray}
I_i &=& \epsilon_{ijk} \big(\langle u_j \frac{\p u_k}{\p x_p} \rangle
    - \frac{1}{\mu \varrho}
    \langle b_j \frac{\p b_k}{\p x_p} \rangle \big)\mB_p
    - \epsilon_{ijk} \big( \langle u_j u_q \rangle \delta_{kp}
    - \frac{1}{\mu \varrho} (\langle b_k b_q \rangle \delta_{jp}
    + 2 \, \langle \frac{\p^2 {\tilde{b}}_q}{\p x_j \p x_p} \, b_k
    \rangle ) \big)
    \frac{\p\mB_p}{\p x_q}
\nonumber\\
Y_i &=& \epsilon_{ijk} \langle \nu (\Delta u_j ) \, b_k + \eta u_j \Delta b_k \rangle
    + \epsilon_{ijk} \langle {\tilde{f}}_j \, b_k \rangle
\label{eq315}\\
Z_i &=& \epsilon_{ijk} \langle {\tilde{T}}_ j \, b_k  + u_j \, G_k \rangle \, .
\nonumber
\end{eqnarray}
The term $\bY$ depends on $\nu$ and $\eta$.
Even if these quantities vanish it remains non--zero as long as
$\epsilon_{ijk} \langle {\tilde{f}}_j \, b_k \rangle$ does so.
The term $\bZ$ considers the non--linear terms $\bG$ and $\tilde{\bT}$ that occur
in (\ref{eq107}) and (\ref{eq141}).
It vanishes if these terms do so.

Let us now split $\bscE$, $\bI$, $\bY$ and $\bZ$ into parts $\bscE^{(0)}$, $\bI^{(0)}$,
$\bY^{(0)}$ and $\bZ^{(0)}$, which are independent of $\bmB$,
and remaining parts $\bscE^{(B)}$, $\bI^{(B)}$, $\bY^{(B)}$ and $\bZ^{(B)}$.
Clearly $\bI^{(0)} = \bzo$, that is $\bI^{(B)} = \bI$.
We do not deal with $\bscE^{(0)}$ and remark only
that it is equal to zero for several simple cases.
For $\bscE^{(B)}$ we have
\begin{equation}
\p_t \bscE^{(B)} = \bI + \bY^{(B)} + \bZ^{(B)} \, .
\label{eq316}
\end{equation}
We introduce now the assumptions
\begin{equation}
\bY^{(B)} = - \bscE^{(B)} / \tau_Y \, , \quad \bZ^{(B)} = - \bscE^{(B)} / \tau_Z
\label{eq317}
\end{equation}
with two times $\tau_Y$ and $\tau_Z$, which are considered as non--negative.
The second of these equations introduces some kind of $\tau$--approximation,
which reduces terms of third order in $\bu$ or $\bbb$ to such of second order.

When accepting (\ref{eq317}) we may rewrite (\ref{eq316}) into
\begin{equation}
\p_t \bscE^{(B)} + (1/\tau_*) \bscE^{(B)}  = \bI \, , \quad 1 / \tau_* = 1 / \tau_Y + 1 / \tau_Z \, .
\label{eq319}
\end{equation}
This is equivalent to
\begin{equation}
\bscE^{(B)} (t) =  \bscE^{(B)} (t_0) \, \exp( - (t - t_0) / \tau_*)
     + \int_0^{t-t_0} \bI (t - t') \, \exp( - t'/ \tau_*) \, \dd t' \, ,
\label{eq321}
\end{equation}
where $t_0$ means some initial time.
We let $t_0 \to -\infty$ and assume that $\bI$ does not markedly vary in time intervals
with a length of some $\tau_*$.
In this way we find a result for $\bscE$ which we can write in the form of (\ref{eq111}),
that is ${\cE}_i = {\cE}^{(0)}_i + a_{ip} \mB_p + b_{ipq} \p \mB_p / \p x_q$, with
\begin{eqnarray}
a_{ip} &=& \epsilon_{ijk} \big(\langle u_j \frac{\p u_k}{\p x_p} \rangle
    - \frac{1}{\mu \varrho}
    \langle b_j \frac{\p b_k}{\p x_p} \rangle \big) \, \tau_*
\nonumber\\
b_{ipq} &=& - \epsilon_{ijk} \big( \langle u_j u_q \rangle \delta_{kp}
    - \frac{1}{\mu \varrho} (\langle b_k \, b_q \rangle \delta_{jp}
    + 2 \, \langle \frac{\p^2 {\tilde{b}}_q}{\p x_j \p x_p} b_k
    \rangle ) \big) \, \tau_* \,.
\label{eq323}
\end{eqnarray}

Let us again consider the special case of a homogeneous isotropic turbulence.
Then $\bscE^{(0)}$ has to be equal to zero
and we have again $a_{ip} = \alpha \, \delta_{ip}$ and $b_{ipq} = \beta \, \epsilon_{ipq}$.
Consequently $\bscE$ takes the form (\ref{eq113}), that is $\bscE = \alpha \, \bmB - \beta \, \bnab \x \bmB$.
With (\ref{eq323}) and considering $\nabla^2 \tilde{\bbb} = - \bbb$ we find
\begin{equation}
\alpha = - \frac{1}{3} \big( \langle \bu \cdot (\bnab \x \bu) \rangle
    - \frac{1}{\mu \varrho} \langle \bbb \cdot (\bnab \x \bbb) \rangle \big) \, \tau_* \, ,
    \quad \beta = \frac{1}{3} \langle u^2 \rangle \, \tau_* \, .
\label{eq325}
\end{equation}
There is a kinetic and a magnetic contribution to $\alpha$ but again only a kinetic one to $\beta$.

It is of some interest to consider also the case of a homogeneous turbulence
deviating from isotropy only due to the presence of a homogeneous mean magnetic field $\bmB$.
Then we have $a_{ip} = \alpha_1 \, \delta_{ip} + \alpha_2 e_{\mathrm{B} i} e_{\mathrm{B} p}$,
where $\be_\mathrm{B} = \bmB / |\bmB|$.
This leads to $\bscE = (\alpha_1 + \alpha_2 ) \bmB$
and $\alpha_1 + \alpha_2 = a_{ip} e_{\mathrm{B} i} e_{\mathrm{B} p}$.
We may write again $\bscE = \alpha \bmB$.
However $\alpha$ has then no longer the form given by (\ref{eq325}) but
\begin{equation}
\alpha = \big( \langle \bu \x (\be_\mathrm{B} \cdot \bnab)\, \bu \rangle \cdot \be_\mathrm{B}
     - \frac{1}{\mu \varrho} \langle \bbb \x (\be_\mathrm{B} \cdot \bnab)\, \bbb \rangle
     \cdot \be_\mathrm{B} \big) \, \tau_* \, .
\label{eq327}
\end{equation}
It is in any case the dependence of $\bu$ and $\bbb$ on $\bmB$
which leads to a dependence of $\alpha$ on $|\bmB|$.
In addition a dependence of $\tau_*$ on $|\bmB|$ cannot be ruled out.

In the limit of small $\bmB$ the averages $\langle \ldots \rangle$ in (\ref{eq323}) and (\ref{eq325})
turn into $\langle \ldots \rangle^{(0)}$, what means that $\bu$ and $\bbb$ are replaced
by the corresponding quantities $\bu^{(0)}$ and $\bbb^{(0)}$ for $\bmB \to \bzo$.
In view of this limit we further remark that the average of the three values of $\alpha$
given by (\ref{eq327}) with $\be_\mathrm{B} = (1, 0, 0)$, $\be_\mathrm{B} = (0, 1, 0)$
and $\be_\mathrm{B} = (0, 0, 1)$ agrees just with $\alpha$ given by (\ref{eq325}),
again with $\bu$ and $\bbb$ replaced by $\bu^{(0)}$ and $\bbb^{(0)}$.

\subsubsection{Comments}
\label{subsubseciia2}

In the above derivation of the result (\ref{eq323}) the assumptions (\ref{eq317})
play a crucial role.
If we simplify our reasoning by considering (\ref{eq319}) from the very beginning
for the steady case only, a very strange aspect of the way to a relation of type (\ref{eq111}),
${\cE}_i = \ldots + a_{ip} \mB_p + b_{ipq} \p \mB_p / \p x_q$, becomes obvious.
There is no original relation of this type, which then would be improved by the assumptions (\ref{eq317}).
Instead, this relation is just established by the assumptions (\ref{eq317}).
$\bscE$ on its left-hand side originates only from the $\bscE^{(B)}$ in (\ref{eq317}).

Nevertheless, at the first glance the assumptions (\ref{eq317}) look plausible at least for simple cases.
For example, for an isotropic background turbulence and a homogeneous steady mean magnetic field $\bmB$
the quantities $\bY^{(B)}$, $\bZ^{(B)}$ and $\bscE^{(B)}$ must by symmetry reasons be proportional to $\bmB$.
This leads immediately to relations like (\ref{eq317}).
The minus signs in these relations together with positive values $\tau_Y$ and $\tau_Z$
ensures that $\bscE$ remains finite.

The assumptions (\ref{eq317}) become however questionable in more complex cases.
Consider, for example, again an isotropic background turbulence but admit a inhomogeneous steady field $\bmB$.
Clearly $\bY^{(B)}$, $\bZ^{(B)}$ and ${\bscE}^{(B)}$ are determined
by the vectors $\bmB$ and $\bnab \x \bmB$ and must have the structures
$c_1 \, \bmB + c_2 \, \bnab \x \bmB$ with two scalar coefficients $c_\nu$, $\nu = 1, 2$.
However, assumptions (\ref{eq317}) could only be justified if the ratios $c_\nu^{(Y)} / c_\nu^{(\cE)}$
and $c_\nu^{(Z)} / c_\nu^{(\cE)}$ of the coefficients $c_\nu^{(Y)}$, $c_\nu^{(Z)}$ and $c_\nu^{(\cE)}$
for $\bY^{(B)}$, $\bZ^{(B)}$ and ${\bscE}^{(B)}$, respectively, were independent of $\nu$.
We see no reason for that.
Some way--out could consist in splitting $\bY^{(B)}$, $\bZ^{(B)}$ and ${\bscE}^{(B)}$ in terms
which correspond either to $c_1 \, \bmB$ or to $c_2 \, \bnab \x \bmB$
and to formulate ansatzes like (\ref{eq317}) for each term separately.

Even if we accept the assumptions (\ref{eq317}) the question remains about realistic values
of $\tau_Y$ and $\tau_Z$, that is, which value of $\tau_*$ should be inserted in the results.

Let us point out a conflict of the results of approach (iia) with those of approach (i).
Consider the special case in which the nonlinear terms $\bG$ and $\bT$ tend to zero,
that is, $\bZ \to \bzo$, or $\tau_Z \to \infty$.
We have to require that then results of approach (iia) like (\ref{eq323}) or (\ref{eq325})
turn into the corresponding results of approach (i).
There is indeed some similarity of the results (\ref{eq323}) and (\ref{eq325})
with the results (\ref{eq227})  and (\ref{eq243}),
or (\ref{eq289}) and (\ref{eq303}) of approach (i) in the limit $q, p \to \infty$ .
However, (\ref{eq323}) and (\ref{eq325}) do not reflect the fact
indicated by (\ref{eq227}) and (\ref{eq289}), or (\ref{eq241}) and (\ref{eq301}),
that $a_{ij}$ and $b_{ijk}$, or $\alpha$ and $\beta$, in this limit depend on correlations
between the components of $\bu$ at two different times.
According to the above reasoning the results (\ref{eq323}) and (\ref{eq325})
should at the same time correspond to the results of approach (i)
for the limit $q, p \to 0$,
that is, (\ref{eq231}) and (\ref{eq247}), or (\ref{eq291}).
This applies obviously at best in a very crude sense.
In particular, the dependence of $a_{ij}$ and $b_{ijk}$, or $\alpha$ and $\beta$,
on correlations of components of $\bu$ in different points in space is not reproduced.

There is another remarkable difference of the results of approaches (iia) and (i).
Consider the case of a purely hydrodynamic background turbulence.
In approach (i) the quantities $a_{ij}$ and $b_{ijk}$
as given by (\ref{eq213}), (\ref{eq227}) or (\ref{eq231})
depend only on the correlation properties
of $\bu$ but, even if $\bu$ is influenced by the magnetic field, not explicitly on those of $\bbb$;
see also section~\ref{subsubsec324}.
In the relation (\ref{eq323}) of approach (iia), however,
in addition to terms with $\bu$ also such with $\bbb$ occur,
which need not to be negligible beyond the limit of small $\bmB$.
In the relations for $\alpha$ given with (\ref{eq325}) and (\ref{eq327}) terms with $\bbb$ occur, too.
Whereas (\ref{eq325}) applies to isotropic turbulence and thus seems reasonable for small $\bmB$ only,
(\ref{eq327}) can be considered for larger $\bmB$, where this term needs no longer to be small.

Admitting again magnetohydrodynamic background turbulence we further note that the times
$\tau^{(\alpha \, u)}$ and $\tau^{(\alpha \, b)}$ in relation (\ref{eq303}) for $\alpha$,
which has been derived in approach (i),
need not to coincide.
In the corresponding result (\ref{eq325}) of approach (iia), however,
only one comparable time, $\tau_*$, occurs.

Since there is hardly any reason to doubt in the results of approach (i)
we conclude that those of approach (iia) are, at least in the range of the validity
of approach (i), not completely correct.
This, of course, calls also results of approach (iia) for more general cases into question.

We further remark that there is no straightforward extension of approach (iia) as presented above
to the case of turbulence in a rotating system.
In this case the equation that occurs instead of (\ref{eq313}) contains in addition to $\bscE$,
which is defined by a product of $\bu$ and $\bbb$ with both taken at the same point,
also a term with a product of $\bu$ and $\bbb$ taken at different points.
Then $\bscE$ can no longer be determined on the simple way used above.
It is to be expected that $a_{ij}$ and $b_{ijk}$ contain correlations
of components of $\bu$ at different points in space.

Finally, even an extension in that sense can hardly deliver a correct result
for a contribution to $\bscE$ like $- \delta_1 (\bOmega \cdot \bnab) \bmB$ in (\ref{eq115}),
because of $(\bOmega \cdot \bnab) \bmB = - \bOmega \x (\bnab \x \bmB) + \bnab (\bOmega \cdot \bmB)$
often discussed as ``$\bOmega \x \bJ$--effect".
In approach (i) this contribution results, at least in the case of homogeneous turbulence,
from a part of the correlation tensor $Q_{ij}$ of $\bu$
which is odd in the difference $\tau$ of the times at which the components of $\bu$ are taken,
see appendix~\ref{app3}.
This part, however, does not occur at all in  approach (iia),
which considers only correlations of the components of $\bu$ at the same time.

The relation (\ref{eq325}) for $\alpha$ is often used as a starting point for investigations
on $\alpha$--quenching, e.g., in \citet{blackmanetal02b} and in \citet{brandenburgetal05,brandenburgetal05b}.
We point out that this relation applies only for isotropic turbulence, that is,
in the limit $\bmB \to \bzo$.
Investigations on $\alpha$--quenching carried out within the framework of approach (iia)
should rather start from a relation like (\ref{eq327}).
On this level, of course, instead of the simple relations for $\langle \bbb \cdot (\bnab \x \bbb) \rangle$
used in the mentioned investigations, which have been concluded from the magnetic helicity balance,
corresponding relations for quantities like
$\langle \bbb \x (\be_\mathrm{B} \cdot \bnab)\, \bbb \rangle \cdot \be_\mathrm{B}$
are needed.
Moreover, as we know from the above comparison with approach (i),
in the case of a purely hydrodynamic background turbulence
the existence of a magnetic contribution to $\alpha$ and so its role in $\alpha$--quenching
is questionable.

\subsection{A spectral $\tau$--approach (approach (iib))}
\label{subseciib}

Let us consider the approach used in the papers by \citet{raedleretal03}
and by \citet{rogachevskiietal03,rogachevskiietal04}.
It has been extensively repeated in the survey article by \citet{brandenburgetal05},
who name it ``minimal $\tau$--approximation (MTA)".
As mentioned above we exclude here again any mean motion and restrict ourselves to a non--rotating fluid,
that is $\bmU = \bOmega = \bzo$.

\subsubsection{Outline of approach (iib)}
\label{subsubseciib1}

In view of the determination of $\bscE$ the attention is first focused on the cross--correlation tensor
$\chi_{jk}$ defined by
\begin{equation}
\chi_{ij} (\bx, \bxi) = \langle u_i (\bx + \bxi/2) \, b_j (\bx - \bxi/2) \rangle \, .
\label{eq331}
\end{equation}
Here and in what follows all quantities are considered at the same time.
For the sake of simplicity we drop the arguments $t$ without ignoring the dependence on $t$.

In addition to $\chi_{ij}$ other correlation tensors are of interest,
which we first define in the general form
\begin{equation}
\Phi_{ij} (\bv, \bw; \bx, \bxi) = \langle v_i (\bx + \bxi/2) \, w_j (\bx - \bxi/2) \rangle \, .
\label{eq333}
\end{equation}
We use a Fourier transformation with respect to the space coordinates only,
$F (\bx) = \int \hat{F} (\bk) \exp( \iu \bk \cdot \bx ) \, \dd^3 k$.
As explained in appendix~\ref{app4} the Fourier transform ${\hat{\Phi}}_{ij}$ of $\Phi_{ij}$
with respect to $\bxi$ can be represented in the form
\begin{equation}
{\hat{\Phi}}_{ij} (\bv, \bw; \bx, \bk)
     =  \int  \langle {\hat{v}}_i (\bk + \bK/2)  \, {\hat{w}}_j (- \bk + \bK/2) \rangle \,
     \exp( \iu \bK \cdot \bx ) \, \dd^3 K \, .
\label{eq337}
\end{equation}

For the calculation of $\bscE$ it is sufficient to know the antisymmetric part of the tensor $\chi_{ij}$,
that is, the vector $\bchi$ defined by $\chi_i = \epsilon_{ijk} \, \chi_{jk}$.
We introduce its Fourier transform with respect to $\bxi$,
\begin{equation}
{\hat{\chi}}_i (\bx, \bk) = \epsilon_{ijk} \, {\hat{\chi}}_{jk} (\bx, \bk) \, .
\label{eq339}
\end{equation}
Then we have
\begin{equation}
\bscE (\bx) = \int \hat{\bchi} (\bx, \bk) \, \dd^3k \, .
\label{eq340}
\end{equation}

Let us consider the time derivative of ${\hat{\chi}}_{ij}$,
\begin{equation}
\p_t {\hat{\chi}}_{ij} (\bx, \bk)
     =  \int  \langle \p_t \hu_i (\bk + \bK/2)  \, \hb_j (- \bk + \bK/2)
     + \hu_i (\bk + \bK/2)  \, \p_t \hb_j (- \bk + \bK/2) \rangle \,
     \exp(\iu \bK \cdot \bx) \, \dd^3 K \, .
\label{eq341}
\end{equation}
We consider $\bscE$ and the quantities necessary for its determination
at a given point and choose the coordinate system such that there $\bx = \bzo$.
In that sense we assume for the calculation of $\p_t \hu_i$ and $\p_t \hb_j$ simply
\begin{equation}
\mB_i (\bx, t) = B_i + B_{ij} \, x_j
\label{eq343}
\end{equation}
with constant $B_i$ and $B_{ij}$.
Then we obtain with the equations (\ref{eq107}), (\ref{eq141}) and (\ref{eq143}),
reduced to $\bOmega = \bmU = \bzo$,
\begin{eqnarray}
\p_t \hu_j (\bk) &=&  - \nu k^2 \hu_j (\bk) + \frac{1}{\mu \varrho} \Big( \iu B_l k_l \hb_j (\bk)
     + B_{jl} \hb_l (\bk) - 2 B_{lm} \frac{k_j k_l}{k^2} \hb_m (\bk)
     - B_{lm} k_l \frac{\p \hb_j (\bk)}{\p k_m} \Big)
\nonumber\\
&& \qquad \qquad \qquad
     + {\hat{\tilde{f}}}_j (\bk) + {\hat{\tilde{T}}}_j (\bk)
\nonumber\\
\p_t \hb_k (\bk) &=&  - \eta k^2 \hb_k (\bk) + \iu B_l k_l \hu_k (\bk) - B_{kl} \hu_l (\bk)
     - B_{lm} k_l \frac{\p \hu_k (\bk)}{\p k_m} + {\hat{G}}_k (\bk) \, .
\label{eq345}
\end{eqnarray}
We insert this into the integral (\ref{eq341}) and split it then into a sum of integrals of the type
\begin{equation}
\int F (\bk, \bK) \, \langle \hu_j (\bk + \bK/2) \, \hu_k (\bk - \bK/2) \rangle \,
     \exp(\iu \bK \cdot \bx) \, \dd^3 K \, .
\label{eq346}
\end{equation}
We may expand $F$ in a power series with respect to $\bK$ and rewrite each factor $K_l$ into a derivative
$\nabla_l = \p / \p x_l$ of the integral.
Since we intend to ignore all contributions to $\bscE$ with higher than first--order spatial derivatives
of mean quantities we do the same with respect to ${\hat{\chi}}_{ij}$.
In this context, of course, $B_{il}$ has to be considered as derivative of the mean quantity $\mB_i$.
Using then (\ref{eq339}) we obtain
\begin{equation}
\p_t \hat{\bchi} =
    - (\nu + \eta) k^2 \hat{\bchi}  + \iu (\nu - \eta) (\bk \cdot \bnab) \, \hat{\bchi}
    + \hat{\bI} + \hat{\bY} + \hat{\bZ} \, .
\label{eq347}
\end{equation}
with
\begin{eqnarray}
{\hat{I}}_i &=& \epsilon_{ijk} \,
    \bigg(- \iu B_l k_l \hu_{jk} (\bx, \bk) + \frac{1}{2} B_l \nabla_l \hu_{jk} (\bx, \bk)
    - B_{kl} \hu_{jl} (\bx, \bk)
\nonumber\\
&& - \frac{1}{2} B_{lm} k_l \frac{\p \hu_{jk} (\bx, \bk)}{\p k_m}
    - \iu B_{lm} k_l x_m \hu_{jk} (\bx, \bk)
\nonumber\\
&& + \frac{1}{\mu \varrho} \Big( \iu B_l k_l \hb_{jk} (\bx, \bk)
     + \frac{1}{2} B_l \nabla_l \, \hb_{jk} (\bx, \bk) + B_{jl} \hb_{lk} (\bx, \bk)
     - 2 B_{lm} \frac{k_j k_l}{k^2} \hb_{mk} (\bx, \bk)
\nonumber\\
&& - \frac{1}{2} B_{lm} k_l \frac{\p \hb_{jk} (\bx, \bk)}{\p k_m}
     + \iu B_{lm} k_l x_m \hb_{jk} (\bx, \bk) \Big) \bigg)
\label{eq349}\\
{\hat{Y}}_i &=& \epsilon_{ijk} \,{\hat{\Phi}}_{jk} (\tilde{\bff}, \bbb; \bx, \bk)
\nonumber\\
{\hat{Z}}_i &=& \epsilon_{ijk} \, \big( {\hat{\Phi}}_{jk} (\tilde{\bT}, \bbb; \bx, \bk)
     + {\hat{\Phi}}_{jk} (\bu, \bG; \bx, \bk) \big) \, .
\nonumber
\end{eqnarray}
The $\hu_{jk}$ and $\hb_{jk}$ are the Fourier transforms of $u_{jk}$ and $b_{jk}$
defined in the sense of (\ref{eq333}) by
\begin{equation}
u_{jk} (\bx, \bxi) = \Phi_{jk} (\bu, \bu; \bx, \bxi) \, , \quad
    b_{jk} (\bx, \bxi) = \Phi_{jk} (\bbb, \bbb; \bx, \bxi) \, .
\label{eq351}
\end{equation}
In contrast to $\bY$ used in approach (iia), $\hat{\bY}$ no longer
contains the diffusion terms, that is neither $\nu$ nor $\eta$,
but is only a consequence of the forcing term in the momentum balance.
Like $\bZ$ in approach (iia) also ${\hat{\bZ}}$ describes the influence
of triple correlations of $\bu$ and $\bbb$ on $\bscE$.

We split now $\bscE$ and likewise $\hat{\bchi}$, $\hat{\bI}$, $\hat{\bY}$ and $\hat{\bZ}$
into parts $\bscE^{(0)}$, ${\hat{\bchi}}^{(0)}$, ${\hat{\bI}}^{(0)}$, ${\hat{\bY}}^{(0)}$
and ${\hat{\bZ}}^{(0)}$ independent of $\bmB$
and remaining parts $\bscE^{(B)}$, ${\hat{\bchi}}^{(B)}$, ${\hat{\bI}}^{(B)}$,
${\hat{\bY}}^{(B)}$ and ${\hat{\bZ}}^{(B)}$.
We have, of course, ${\hat{\bI}}^{(0)} = \bzo$, that is, ${\hat{\bI}}^{(B)} = {\hat{\bI}}$.
Again we  do not deal with ${\cE}^{(0)}$.
In view of $\bscE^{(B)}$ we conclude from (\ref{eq347}) that
\begin{equation}
\p_t {\hat{\bchi}}^{(B)} =
    - (\nu + \eta) k^2 {\hat{\bchi}}^{(B)} + \iu (\nu - \eta) (\bk \cdot \bnab) \, {\hat{\chi}}^{(B)}
    + \hat{\bI} + {\hat{\bY}}^{(B)} + {\hat{\bZ}}^{(B)} \, .
\label{eq353}
\end{equation}

In analogy to (\ref{eq317}) we introduce here the assumptions
\begin{equation}
{\hat{\bY}}^{(B)} (\bx, \bk) = - {\hat{\bchi}}^{(B)} (\bx, \bk) / {\hat{\tau}}_Y (k) \, , \quad
     {\hat{\bZ}}^{(B)} (\bx, \bk) = - {\hat{\bchi}}^{(B)} (\bx, \bk) / {\hat{\tau}}_z (k)
\label{eq363}
\end{equation}
with positive functions ${\hat{\tau}}_Y$ and ${\hat{\tau}}_Z$ of $k$.
In \citet{raedleretal03} it has not been distinguished
between the two terms ${\hat{\bY}}^{(B)}$ and ${\hat{\bZ}}^{(B)}$,
although they are of quite different origin.
For their sum an ansatz like that for ${\hat{\bY}}^{(B)}$ or ${\hat{\bZ}}^{(B)}$ is used.
This is, however, of minor interest for the following discussions.

Starting from (\ref{eq353}) and using (\ref{eq363}) we obtain
\begin{equation}
\p_t {\hat{\bchi}}^{(B)} + (1/{\hat{\tau}}_*) {\hat{\bchi}}^{(B)}
   = \iu \, (\nu - \eta) (\bk \cdot \bnab) \, {\hat{\bchi}}^{(B)}
   + \hat{\bI} \, , \quad
   1/{\hat{\tau}}_* = (\nu + \eta) \, k^2 + 1/{\hat{\tau}}_Y + 1/{\hat{\tau}}_Z \, .
\label{eq367}
\end{equation}
We recall that all contributions to $\bscE$ with higher than first--order derivatives
of mean quantities should be ignored.
This applies, of course, to ${\hat{\bchi}}^{(B)}$ and $\hat{\bI}$, too.
In that sense we conclude from (\ref{eq367}) that
\begin{equation}
\p_t (\bk \cdot \bnab) \, {\hat{\bchi}}^{(B)} + (1/{\hat{\tau}}_*) (\bk \cdot \bnab) \, {\hat{\bchi}}^{(B)}
   = (\bk \cdot \bnab) \, {\hat{\bI}} \, .
\label{eq369}
\end{equation}
In view of $\bscE$  at $\bx = \bzo$ we consider also the equations (\ref{eq367}) and (\ref{eq369})
at $\bx = \bzo$.
They are then differential equations for ${\hat{\bchi}}^{(B)}$ and $(\bk \cdot \bnab) {\hat{\bchi}}^{(B)}$
with respect to $t$.
Solving them in the same way as we did it with (\ref{eq319}),
assuming here, too, $t_0 \to - \infty$ and sufficiently small variations of $\hat{\bI}$ in time,
we find
\begin{equation}
{\hat{\bchi}}^{(B)} = \big( \hat{\bI} + \iu (\nu - \eta) (\bk \cdot \bnab) \hat{\bI} \, {\hat{\tau}}_*\big) \, {\hat{\tau}}_* \, ,
\label{eq371}
\end{equation}
and, together with (\ref{eq349}),
\begin{eqnarray}
{\hat{\chi}}^{(B)}_i &=& \epsilon_{ijk} \,
    \bigg( \Big(\frac{1}{2} \nabla_p \, \hu_{jk}
    -  \iu k_p \hu_{jk}
    + \frac{1}{\mu \varrho}
    ( \frac{1}{2} \nabla_p \, \hb_{jk}
    + \iu k_p \hb_{jk} )
\nonumber\\
&& \quad \quad \quad
    + (\nu - \eta) \, k_l k_p \nabla_l \, ( \hu_{jk}
    - \frac{1}{\mu \varrho} \, \hb_{jk} ) \, {\hat{\tau}}_* (k) \Big) \,
    {\hat{\tau}}_* (k) \, B_p
\nonumber\\
&& - \Big( \hu_{jq} \delta_{kp}
    + \frac{1}{2} \, k_p \, \frac{\p}{\p k_q} \hu_{jk}
    - \frac{1}{\mu \varrho} ( \hb_{qk} \delta_{jp}
    - \frac{1}{2} \, k_p \, \frac{\p}{\p k_q} \hb_{jk}
    - 2 \, \frac{k_j k_p}{k^2} \, \hb_{qk} )
\label{eq373}\\
&& \quad \quad \quad
    - (\nu - \eta) \, k_p k_q \big( \hu_{jk}
    - \frac{1}{\mu \varrho} \, \hb_{jk} \big) \, {\hat{\tau}}_* (k) \Big) \,
    {\hat{\tau}}_* (k) B_{pq} \bigg) \, ,
\nonumber
\end{eqnarray}
again with all quantities taken at $\bx = \bzo$.

With this result we can calculate $\bscE$ according to (\ref{eq340}).
Admitting again arbitrary $\bx$ and writing the result in the form (\ref{eq111}),
that is, ${\cE}_i = {\cE}^{(0)}_i + a_{ij} \mB_j + b_{ijk} \p \mB_j / \p x_k$, we have
\begin{eqnarray}
a_{ip} &=& \epsilon_{ijk} \, \int
    \bigg( \frac{1}{2} \nabla_p \, \hu_{jk} (\bx, \bk) -  \iu k_p \hu_{jk} (\bx, \bk)
    + \frac{1}{\mu \varrho}
    \Big( \frac{1}{2} \nabla_p \, \hb_{jk} (\bx, \bk) + \iu k_p \hb_{jk} (\bx, \bk)\Big)
\nonumber\\
&& \quad \quad \quad
    + (\nu - \eta) \, k_l k_p \nabla_l \, \Big( \hu_{jk} (\bx, \bk)
    - \frac{1}{\mu \varrho} \, \hb_{jk} (\bx, \bk)\Big) \, {\hat{\tau}}_* (k) \bigg) \,
    {\hat{\tau}}_* (k) \, \dd^3 k
\nonumber\\
b_{ipq} &=& - \epsilon_{ijk} \int
    \bigg( \hu_{jq} (\bx, \bk)\delta_{kp}
    + \frac{1}{2} \, k_p \, \frac{\p}{\p k_q} \hu_{jk} (\bx, \bk)
\label{eq377}\\
&& \quad \quad \quad
    - \frac{1}{\mu \varrho} \Big( \hb_{qk} (\bx, \bk)\delta_{jp}
    - \frac{1}{2} \, k_p \, \frac{\p}{\p k_q} \hb_{jk} (\bx, \bk)
    - 2 \, \frac{k_j k_p}{k^2} \, \hb_{qk} (\bx, \bk)\Big)
\nonumber\\
&& \quad \quad \quad
    - (\nu - \eta) \, k_p k_q \Big( \hu_{jk} (\bx, \bk)
    - \frac{1}{\mu \varrho} \, \hb_{jk} (\bx, \bk)\Big) \, {\hat{\tau}}_* (k) \bigg) \,
    {\hat{\tau}}_* (k) \, \dd^3k \, .
\nonumber
\end{eqnarray}
These relations are by the convolution theorem equivalent to
\begin{eqnarray}
a_{ip} &=& \frac{\epsilon_{ijk}}{(2 \pi)^3}
    \bigg( \int_\infty \Big( \langle u_j (\bx + \bxi/2, t)
    \frac{\p u_k (\bx - \bxi/2,t)}{\p x_p} \rangle \,
\nonumber\\
&& \quad \quad \quad
    + \frac{1}{\mu \varrho} \,
    \langle \frac{\p b_j (\bx + \bxi/2, t)}{\p x_p}
    b_k (\bx - \bxi/2, t)\rangle \Big)\,
    \tau_* (\xi) \, \dd^3 \xi
\nonumber\\
&& \quad \quad \quad
    - (\nu - \eta) \, \frac{\p}{\p x_l} \,
    \int_\infty \frac{\p^2}{\p \xi_l \p \xi_p}
    \Big( \langle u_j (\bx + \bxi/2, t)\, u_k (\bx - \bxi/2, t)\rangle
\nonumber\\
&& \quad \quad \quad \quad \quad \quad \quad \quad \quad
    - \frac{1}{\mu \varrho} \,
    \langle b_j (\bx + \bxi/2, t)\, b_k (\bx - \bxi/2, t)\rangle \Big) \,
    {\breve{\tau}}_*^2 (\xi) \, \dd^3 \xi \, \bigg)
\nonumber\\
b_{ipq} &=& - \frac{\epsilon_{ijk}}{(2 \pi)^3}
    \Bigg( \int_\infty \bigg(
    \langle u_j (\bx + \bxi/2, t)\, u_q (\bx - \bxi/2, t)\rangle \, \delta_{kp}
\nonumber\\
&& \qquad \qquad \qquad
    - \frac{\xi_q}{2} \frac{\p}{\p \xi_p} \,
    \langle u_j (\bx + \bxi/2, t)\, u_k (\bx - \bxi/2, t)\rangle
\label{eq381}\\
&& \quad \quad \quad
    - \frac{1}{\mu \varrho} \,
    \Big( \langle b_q (\bx + \bxi/2, t)\, b_k (\bx - \bxi/2, t)\rangle \,
    \delta_{jp}
\nonumber\\
&& \qquad \qquad \qquad
    + \frac{\xi_q}{2} \frac{\p}{\p \xi_p} \,
    \langle b_j (\bx + \bxi/2, t)\, b_k (\bx - \bxi/2, t)\rangle
\nonumber\\
&& \quad \quad \quad \quad \quad \quad
    - 2 \, \frac{\p^2}{\p \xi_j \p \xi_p}
    \big( \Delta_\xi^{-1} \langle b_q (\bx + \bxi/2, t)b_k (\bx - \bxi/2, t)\rangle \big) \Big) \bigg)
    \tau_* (\xi) \, \dd^3 \xi
\nonumber\\
&& \quad \quad \quad
    + (\nu - \eta) \, \int_\infty
    \frac{\p^2}{\p \xi_q \p \xi_p}
    \Big( \langle u_j (\bx + \bxi/2, t)\, u_k (\bx - \bxi/2, t)\rangle
\nonumber\\
&& \quad \quad \quad \quad \quad \quad \quad \quad \quad
    - \frac{1}{\mu \varrho}
    \langle b_j (\bx + \bxi/2, t)\, b_k (\bx - \bxi/2, t)\rangle \Big) \,
    {\breve{\tau}}_*^2 (\xi) \, \dd^3 \xi \Bigg) \, .
\nonumber
\end{eqnarray}
For clarity we have inserted again the arguments $t$.
The quantities $\tau_*$ and ${\breve{\tau}}_*$ are given by
\begin{eqnarray}
\tau_* (\xi) &=& \int {\hat{\tau}}_* (k) \, \exp(\iu \bk \cdot \bxi) \, \dd^3 k
    = \frac{4 \pi}{\xi} \int_0^\infty {\hat{\tau}}_* (k) \, \sin (k \xi) \, k \, \dd k
\nonumber\\
{\breve{\tau}}_*^2 (\xi) &=& \int {\hat{\tau}}_*^2 (k) \exp(\iu \bk \cdot \bxi) \, \dd^3 k
    = \frac{1}{(2 \pi)^3} \int_\infty \tau_* (|\bxi - \bxi'|) \, \tau_* (\xi') \,
    \dd^3 \xi' \, .
\label{eq383}
\end{eqnarray}
Note that $\tau_*$ has the dimension $\mbox{time}/{\mbox{length}}^3$,
and ${\breve{\tau}}_*$ the dimension $\mbox{time}/{\mbox{length}}^{3/2}$.
As for $\Delta_\xi^{-1}$ we refer to the definition (\ref{eq283}).

Let us again consider the special case of homogeneous isotropic turbulence.
Then we have again $\bscE^{(0)} = \bzo$
and $a_{ip} = \alpha \, \delta_{ip}$ and $b_{ipq} = \beta \, \epsilon_{ipq}$
so that $\bscE$ takes the form (\ref{eq113}), that is $\bscE = \alpha \, \bmB - \beta \bnab \x \bmB$.
Using (\ref{eq381}) we find
\begin{eqnarray}
\alpha &=& - \frac{1}{3 (2 \pi)^3} \int_\infty
    \Big( \langle \bu (\bx + \bxi/2, t)\cdot \big(\bnab \x \bu (\bx - \bxi/2, t)\big) \rangle
\nonumber\\
&& \quad \quad \quad \quad \quad \quad \quad
    - \frac{1}{\mu \varrho} \,
    \langle \bbb (\bx + \bxi/2, t)\cdot \big(\bnab \x \bbb (\bx - \bxi/2, t)\big) \rangle \Big) \,
    \tau_* (\xi) \, \dd^3 \xi
\nonumber\\
\beta &=& \frac{1}{3 (2 \pi)^3} \int_\infty
    \langle \bu (\bx + \bxi/2, t)\cdot \bu (\bx - \bxi/2, t)\rangle \,
    \tau_* (\xi) \, \dd^3 \xi \, .
\label{eq391}
\end{eqnarray}
We have here replaced
$\langle \bbb (\bx - \bxi/2, t)\cdot \big(\bnab \x \bbb (\bx + \bxi/2, t)\big) \rangle$,
what originally occurs in the derivation from (\ref{eq381}),
by $\langle \bbb (\bx + \bxi/2, t)\cdot \big(\bnab \x \bbb (\bx - \bxi/2, t)\big) \rangle$.
Due to the assumed homogeneity of the turbulence the averages $\langle \ldots \rangle$ in (\ref{eq391})
do not depend on $\bx$ , and we may further replace, e.g.,
$\langle \bu (\bx + \bxi/2, t) \cdot \big(\bnab \x \bu (\bx - \bxi/2, t)\big) \rangle$
with $\langle \bu (\bx, t) \cdot \big(\bnab \x \bu (\bx + \bxi, t)\big) \rangle$.
Again, by the reason explained in the context of (\ref{eq237}) the integrals $\int_\infty \ldots \dd^3 \xi$
may be replaced with $4 \pi \int_0^\infty \ldots \xi^2 \dd \xi$.
The $(\nu - \eta)$ terms in (\ref{eq381}) are without influence
on $\alpha$ and $\beta$.

In \citet{raedleretal03} the possibility of a non--zero $\bscE^{(0)}$ has been ignored
from the very beginning.
In addition only the limit of small $\bmB$ is considered.
So for a comparison $\hu_{ij}$ and $\hb_{ij}$ in (\ref{eq373}) and (\ref{eq377}) have to be interpreted
as $\hu^{(0)}_{ij}$ and $\hb^{(0)}_{ij}$, the limits of these quantities for $\bmB \to \bzo$,
and likewise $\bu$ and $\bbb$ in (\ref{eq381}) and (\ref{eq391}) as $\bu^{(0)}$ and $\bbb^{(0)}$,
understood in the same sense.

\subsubsection{Comments}
\label{seciibc3}

Similar to the situation with approach (iia) and the assumptions (\ref{eq317})
on $\bY^{(B)}$ and  $\bZ^{(B)}$
the results of the above derivation depend crucially on the assumptions (\ref{eq363})
on ${\hat{\bY}}^{(B)}$ and ${\hat{\bZ}}^{(B)}$ describing the influences
of the forcing and of the triple correlations of $\bu$ and $\bbb$.
So we have here essentially  to repeat the comment made in \ref{subsubseciia2}
in view of the assumptions (\ref{eq317}) used in approach (iia).
Here, too, we see no rigorous way to justify assumption (\ref{eq363}).
Moreover, the choice of ${\hat{\tau}}_Y$ and ${\hat{\tau}}_Z$ remains open.

A very strange aspect of the above procedure is that it needs as input no other correlation tensors
than those with components of $\bu$ or $\bbb$ taken at the same time.

Let us again consider the case in which the nonlinear terms $\bG$ and $\bT$ tend to zero.
This implies ${\hat{\bZ}}^{(B)} \to \bzo$, or $\tau_Z \to \infty$.
In this case the results of approach (iib) have to coincide with those of approach (i).

As can be seen, e.g., from (\ref{eq367}) the results of approach (iib) depend in general
on both $\nu$ and $\eta$.
This is clearly in conflict with the fact that those of approach (i) for purely hydrodynamic turbulence
do not depend on $\nu$.

In the limit $q \to \infty$ approach (i) delivered us the quantities $a_{ij}$ and $b_{ijk}$
in the form (\ref{eq227}).
These are far from coinciding with the results (\ref{eq381}) obtained in approach (iib).
A striking discrepancy is that in approach (i), apart from the independence of the results on $\nu$,
only correlations of the components of $\bu$ at the same point in space but at different times occur,
in approach (iib) only correlations at different points in space but at the same time.
At the first glance this discrepancy seems to disappear in the case
of homogeneous isotropic statistical steady turbulence,
in which the results of approach (i) for $\alpha$ and $\beta$ given by (\ref{eq241}) can be rewritten
in the forms (\ref{eq519}) and (\ref{eq523}), which correspond to (\ref{eq391}).
While however the $\tau_c$ (\ref{eq519}) and (\ref{eq523}) may differ from each other
the $\tau_*$ in (\ref{eq241}) have to coincide.

In the limit $q \to 0$ approach (i) delivers us the relations (\ref{eq231})
for $a_{ij}$ and $b_{ijk}$.
They are in so far similar to (\ref{eq381}) as in both cases only correlations
of the components of $\bu$ at the same time occur.
Nevertheless a coincidence of the two kinds of results would require very special properties
of the factor $\tau_*$ in (\ref{eq391}).
We see no reasons for such properties.
When putting on trial ${\hat{\bY}}^{(B)} = \bzo$
(\citet{brandenburgetal05} consider ${\hat{\bY}}^{(B)}$ as negligible)
we arrive at expressions for $\alpha$ and $\beta$
which differ from (\ref{eq247}) only in so far as $\eta + \nu$ occurs instead of $\eta$.
A view at the first of the equations (\ref{eq345}) might suggest
to put ${\hat{\bY}}^{(B)} = \nu k^2 {\hat{\bchi}}^{(B)}$.
Then we obtain indeed $\alpha$ and $\beta$ in full agreement with (\ref{eq247}).
Nevertheless ${\hat{\bY}}^{(B)} = \nu k^2 {\hat{\bchi}}^{(B)}$ removes $\nu$ only from one of two places
at which it occurs in (\ref{eq367}), and even if it does no longer occur in $\alpha$ and $\beta$
it will do so in other mean--field coefficients.
We believe that an essential reason for shortcomings of approach (iib)
is a treatment of the term ${\hat{\bY}}^{(B)}$, which is in general not justified.

In the discussion of the results of approach (iia) we have pointed out a difference
to those of approach (i),
which becomes visible in the special case of purely hydrodynamic background turbulence.
This difference occurs here, too.
In this case in approach (i) the quantities $a_{ij}$ and $b_{ijk}$, or $\alpha$ and $\beta$,
can always be represented by the correlation properties of $\bu$.
In approach (iib) however, beyond the limit of small $\bmB$, the expressions for these quantities
except $\beta$ contain both $\bu$ and $\bbb$.

We see that approach (iib) to $\bscE$ delivers results which even under comparable conditions
deviate from the results of approach (i), which we consider as correct.
Therefore also beyond the range of validity of approach (i) and beyond the case of non--rotating turbulence,
all results derived in the framework of approach (iib) have to be considered with caution.
Their details are in general not correct.

If we admit a Coriolis force, that is $\bOmega \not= \bzo$, or a mean motion, $\bmU \not= \bzo$,
there is a coupling between the symmetric and antisymmetic parts of ${\hat{\chi}}_{ij}$,
and the calculation of $\hat{\bchi}$ is more complex.
By the reason already discussed in section \ref{subsubseciia2} it is then hard to imagine
that an extension of approach (iib) to the case $\bOmega \not= \bzo$ can describe,
e.g., the ``$\bOmega \x \bJ$--effect" correctly.
Here a view on the ``$\bOmega \x \bJ$--dynamo" is of some interest.
Its existence depends on a non--zero value of a specific coefficient.
In \citet{raedleretal03}, equations (130) and (132), it is given by $\delta_0 - \kappa_0$,
and for hydrodynamic background turbulence it is non--zero as long as ${\hat{\tau}}_*$ depends on $k$.
According to the calculations by \citet{raedleretal06}
done again for hydrodynamic background turbulence and in the second--order correlation approximation,
however, this value vanishes in both the limits $q \to 0$ and $q \to \infty$
independent of any assumption on the correlation time;
see section~VD and appendix~C of that paper.

\citet{rogachevskiietal03}, considering a mean motion and proceeding in the spirit of approach (iib),
found a contribution to $\bscE$ proportional to $\bW \x (\bnab \x \bmB)$,
where $\bW$ stands for $\bnab \x \bmU$.
A contribution which can be interpreted in that sense occurs already in a paper by \citet{urpin99},
though as a result of a not easily comprehensible calculation.
The existence of this ``$\bW \x \bJ$--effect" has been confirmed in the framework
of the second--order correlation approximation by \citet{raedleretal06}.
The simple model of an ``$\bW \x \bJ$--dynamo" proposed by \citet{rogachevskiietal03}
(see also Rogachevskii and Kleeorin 2004) works with the coefficients
determining this and the accompanying effects as calculated in the spectral $\tau$--approximation.
It turned out however that it fails to work with the coefficients resulting
from the second--order correlation approximation for purely hydrodynamic background turbulence,
and this applies independently of $q$
\citep{ruedigeretal06,raedleretal06}.

\subsection{The $\tau$--approaches in the light of numerical simulations}
\label{subsec43}

Quite naturally the question arises whether the correctness or the limits of validity
of the $\tau$--approaches can be checked by means of direct numerical simulations.
Indeed, \cite{brandenburgetal05b} report on numerical experiments with
forced helical magnetohydrodynamic turbulence in a rectangular box
aimed at studying the $\alpha$--effect and $\alpha$--quenching in the framework of approach (iia).
In the underlying theory they use the ansatz (\ref{eq317}) for $\bZ^{(B)}$
(in their notation $\ol{\bT}$), standing for triple correlations,
but ignore $\bY^{(B)}$, which comprises the effects of viscous and ohmic dissipation
and the correlation of the magnetic fluctuations with the external force.

Brandenburg and Subramanian examined the relation between $\bZ^{(B)}$
and a dynamo--generated mean field $\bmB$
on the basis of a simulation with a forcing yielding an isotropic background turbulence.
Under the assumption of the proportionality of $\bscE$ and $\bmB$,
justified by a weak spatial variation of $\bmB$,
the result turned out to be consistent with the ansatz (\ref{eq317}) for $\bZ^{(B)}$.
The simulation confirmed not only the proportionality of $\bZ^{(B)}$ and $\bscE$,
which was due to the isotropy of the turbulence to be expected already by symmetry reasons,
but delivered also a non--negative $\tau_Z$.
The ansatz (\ref{eq317}) for $\bY^{(B)}$, which was not considered in this work,
is plausible by symmetry reasons, too.
So under the considered circumstances the neglect of a non--zero $\bY^{(B)}$
cannot have any other consequence in the result for $\alpha$
than a deviation of the time $\tau_*$ from $\tau_Z$.

In a further step Brandenburg and Subramanian
extracted the time $\tau_*$ (in their notation $\tau$)
from the simulations by making use of the relation (\ref{eq325}) for $\alpha$.
Of course, using (\ref{eq325}) instead of (\ref{eq327}) has generally
to be questioned as only the latter is the relevant one beyond the limit $\bmB \to \bzo$.
The use of (\ref{eq325}) can only be justified under the additional assumption that
$- \frac{1}{3} \, \langle \bu \cdot (\bnab \x \bu) \rangle$
and $- \frac{1}{3} \, \langle \bbb \cdot (\bnab \x \bbb) \rangle$
are proper proxies of $\langle \bu \x (\be_\mathrm{B} \cdot \bnab)\, \bu \rangle \cdot \be_\mathrm{B}$
and $\langle \bbb \x (\be_\mathrm{B} \cdot \bnab)\, \bbb \rangle \cdot \be_\mathrm{B}$.
In some cases, from the simulations negative values of $\tau_*$ emerged
(surprisingly enough just for small $\bmB$),
what is in contradiction with the idea of a mean electromotive force
decaying in the absence of a mean magnetic field.
As a way--out an ad-hoc modification of (\ref{eq325}) was introduced
in form of additional quenching functions, $g_K$ and $g_M$,
in front of both the kinetic and the magnetic contributions to $\alpha$.
This measure cannot be brought into accordance
with the derivation of (\ref{eq325}), which allows at best the same quenching function
for both contributions, that is, a dependence of $\tau_*$ on $|\bmB|$.
Interestingly enough, in the relations (\ref{eq297}) and  (\ref{eq303})
for $\alpha$, derived in approach (i) for magnetohydrodynamic background turbulence
in the limit $q, p \to \infty$,
the two times $\tau^{(\alpha \, u)}$ and $\tau^{(\alpha \, b)}$ occur,
which could be well different from each other.
As far as the assumption of magnetohydrodynamic background turbulence,
that is, of a small-scale dynamo, is justified,
the numerical results support our doubt in the correctness of approach (iia)
based on the conflict of the latter with approach (i).
It is however not clear to us whether all runs correspond to this assumption.
In the case of a purely hydrodynamic background turbulence there is still another doubt
in relation (\ref{eq325}) for $\alpha$, which results also from the comparison with approach (i)
and concerns the existence of the magnetic contribution to $\alpha$ (see section~\ref{subsubseciia2}).

Let us for a moment deviate from approach (iia)
and accept the concept employing the two quenching functions $g_K$ and $g_M$.
In order to get enough constraints for their unambiguous determination
results of simulations with different (kinetic and magnetic) forcings were employed.
This seems to be again questionable, as only the structures of relations
like (\ref{eq325}) or (\ref{eq327}) can be considered as universal in their validity ranges,
but not the specific values of $\tau_*$ or of $g_K$ and $g_M$.
These values should in general differ for different realizations of turbulence.
As a consequence, they can at best be determined using data
from one and the same simulation or from simulations which are equivalent
on the level of mean quantities.
Examining the visualizations of the employed realizations of turbulence
given in \citet{brandenburgetal05b} (figures 10 and 11)
it is by no means clear whether this requirement is fulfilled.
But even if so, an unambiguous determination of the two quenching factors were then again impossible
because the relevant equations (17) of that paper would just become linearly dependent.

In view of some support of approach (iia) from numerical studies we mention also the investigations
of \citet{blackmanetal02b} and \citet{fieldetal02} on the saturation of a mean--field dynamo.
Specific results on helicities obtained in this approach,
even if using a relation for $\alpha$ like (\ref{eq325}) instead of one like (\ref{eq327}),
are surprisingly well matched to those of numerical simulations by \citet{brandenburg01b},
which are independent of such an approach.

Summing up, we state that a convincing numerical support of the $\tau$--approach
still requires further efforts beyond the work published so far.

In this context studies of the behavior of passive scalars in a turbulent fluid carried out
by \citet{blackmanetal03} and by \citet{brandenburgetal04} using the $\tau$--approach, too,
deserve some interest.
Numerical simulations confirm the results derived in the framework of this approximation rather well.
This is remarkable since the discussed concerns with respect to the treatment of the kinetic forcing term
(see section~\ref{seciibc3}) apply here, too.

\section{An alternative}
\label{new}

As explained above approach (i) is based on approximative solutions
of the induction equation (\ref{eq107}) for $\bbb$.
The approximation is defined by the restriction to second--order correlations
in the  velocity fluctuations $\bu$.
In contrast to this in the approaches (ii) neither the induction equation nor the momentum equation
are really solved.
Instead simple assumptions are used for some crucial quantities.
These assumptions, however, have proved to be very problematic.
We believe that a real progress in the calculation of the mean electromotive force $\bscE$
beyond the second--order correlation approximation can only be gained on the basis
of improved solutions of the induction equation (\ref{eq107}).
In the expressions for $\bscE$ obtained in this way the quantities depending on $\bu$
can then be specified according to the momentum equation.

\subsection{Higher--order correlation approximations}
\label{new1}

As is well known it is possible to proceed from the second--order correlation approximation
for the mean electromotive force $\bscE$ to higher--order approximations
\citep{krauseetal71b,krauseetal80}.
Basically arbitrarily high orders can be reached.
It has been shown that the proposed procedure converges
\citep{krause68,krauseetal80}.
However, the calculation of $\bscE$ in a higher--order approximation requires the knowledge
of higher--order correlation tensors and is moreover very tedious.
That is why so far only in a few simple cases higher--order results have been given
(e.g., Nicklaus and Stix 1988, Carvalho 1992, 1994, R\"adler {\it et al.} 1997).

In appendix~\ref{app5} a procedure is described for the calculation of $a_{ij}$ and $b_{ijk}$
in higher--order correlation approximations.
For the sake of simplicity it is restricted to cases with $\bmU = \bzo$ and $\bnab \cdot \bu = 0$.
It turns out that then $a_{ij}$ and $b_{ijk}$ can be represented in the simple form
\begin{eqnarray}
a_{ij} (\bx, t) &=&  \epsilon_{ilm}
   \int_0^\infty \!\!\! \int_\infty  \frac{\p G (\xi, \tau)}{\p \xi_j} \,
   \ol{u_l (\bx, t) \, v_m (\bx, t; - \bxi, - \tau)} \, \dd^3 \xi \, \dd \tau \, \quad
\nonumber\\
b_{ijk} (\bx, t) &=&  \epsilon_{ilm}
   \int_0^\infty \!\!\! \int_\infty  G (\xi, \tau) \,
   \ol{u_l (\bx, t) \, w_{mjk} (\bx, t; - \bxi, - \tau)} \, \dd^3 \xi \, \dd \tau \, .
\label{eq392}
\end{eqnarray}
The quantities $v_m$ and $w_{ljk}$ are given as sums of constituents $v^{(\nu)}_m$ and $w^{(\nu)}_{ljk}$,
where $\nu$ is the order in $\bu$.
Recursions for the calculation of $v^{(\nu)}_m$ from $v^{(\nu-1)}_m$
and of $w^{(\nu)}_{ljk}$ from $w^{(\nu-1)}_{ljk}$ are available, see appendix~\ref{app5}.

Let us add a remark concerning $\bscE^{(0)}$.
As explained above, equation (\ref{eq107}) for $\bbb$ in its original form,
that is, without the neglect of $\bG$, with a properly specified $\bu$
may have non--decaying solutions even if $\bmB$ is equal to zero.
Then $\bscE^{(0)}$ may be unequal to zero.
As already noted in section~\ref{subsubsec311} in the second--order correlation approximation,
when ignoring the influence of the initial $\bbb$,
no such solutions $\bbb$ exist and therefore $\bscE^{(0)}$ vanishes.
Hence, if we start from the second--order approximation and proceed then to higher
approximations we never find $\bscE^{(0)}$, though it may well be non--zero.
In that sense, the iteration procedure is not complete.

\subsection{A closure proposal}

Instead of approaches of type (ii) to the mean electromotive force $\bscE$ we propose another one.
For the sake of simplicity we restrict ourselves again to the case $\bmU = \bzo$
and ignore the influence of any initial value of $\bbb$, that is, consider the limit $t_0 \to - \infty$.
Then equation (\ref{eq107}) is equivalent to
\begin{eqnarray}
b_k (\bx, t) &=& (\delta_{km} \delta_{ln} - \delta_{kn} \delta_{lm})
    \int_0^{\infty} \int_\infty \frac{\p G (\bxi, \tau)}{\p \xi_l}
    \Big(u_m (\bx - \bxi, t - \tau) \, \mB_n (\bx -\bxi, t - \tau)\label{eq401}\\
&& \qquad \qquad \qquad \qquad \qquad \qquad \qquad
    + \big(u_m (\bx - \bxi, t - \tau) \, b_n (\bx -\bxi, t - \tau)\big)' \Big) \, \dd^3 \xi \, \dd \tau \, .
\nonumber
\end{eqnarray}
This leads immediately to
\begin{eqnarray}
{\cE}_i (\bx, t) &=& (\delta_{il} \epsilon_{kmn} - \delta_{kl} \epsilon_{imn})
    \int_0^{\infty} \int_\infty \frac{\p G (\bxi, \tau)}{\p \xi_l}
    \Big( \langle u_k (\bx, t) \, u_m (\bx - \bxi, t - \tau) \rangle \, \mB_n (\bx -\bxi, t - \tau)
\label{eq403}\\
&& \qquad \qquad \qquad \qquad \qquad \qquad
    - \langle u_k (\bx, t) \, u_m (\bx - \bxi, t - \tau) \, b_n (\bx -\bxi, t - \tau) \rangle \Big) \,
    \dd^3 \xi \, \dd \tau \, .
\nonumber
\end{eqnarray}
We write for short
\begin{equation}
\bscE = \bscE^{(\mathrm{SOCA})} + \bscE^{(uub)} \, ,
\label{eq405}
\end{equation}
where $\bscE^{(\mathrm{SOCA})}$ is the result for $\bscE$ obtained in the second--order correlation approximation,
that is, with (\ref{eq403}) if there the part of the integrand with $b_n$ is ignored,
and $\bscE^{(uub)}$ the contribution to $\bscE$ defined by this part.

Our proposal for a new approach to $\bscE$ consists in its simplest form in putting
\begin{equation}
\bscE^{(uub)} = - f \bscE^{(\mathrm{SOCA})}
\label{eq407}
\end{equation}
with some factor $f$, which will be discussed below.
Then we have
\begin{equation}
\bscE = (1 - f) \,\bscE^{(\mathrm{SOCA})} \, .
\label{eq409}
\end{equation}

The ansatz (\ref{eq407}) corresponds in a formal sense to the ansatzes (\ref{eq317}) or (\ref{eq363})
used in the approaches (ii).
As the time $\tau_*$ in these cases, the factor $f$ remains at first undetermined.
In order to find rough statements or estimates for the factor $f$ a look on the results
of higher--order approximations reported above is useful.
These results suggest that, e.g., in the limit $q \to 0$ the factor $f$
has the form $f = c_1 Rm$ with some positive $c_1$ for small $Rm$,
and $f = c_1 Rm + c_2 Rm^2 + \ldots$ for higher $Rm$.
In the limit $q \to \infty$ the factor $f$ should have the form $f = c_1 St$ for small $St$,
and the form $f = c_1 St + c_2 St^2 + \ldots$ for higher $St$.
Since $f \to 0$ as $Rm \to 0$ or $St \to 0$, respectively, the ansatz (\ref{eq407})
trivially satisfies the requirement that higher approaches should include
the second--order correlation approximation in overlapping parts
of their respective validity regions.

In this context the $\alpha$--effect calculations for the Karlsruhe experiment are of particular interest
\citep{raedleretal97,raedleretal98,raedleretal02}.
They have been done for the case $\bmU = \bzo$ and steady flows $\bu$, that is $q = 0$,
with arbitrary $Rm$.
The flows have the form $\bu = \bu_\perp + \bu_\parallel$, where $\bu_\perp$ and $\bu_\parallel$
are, with respect to a proper Cartesian coordinate system $(x,y,z)$,
defined by $\bu_\perp = \big( u_x (x,y), u_y (x,y), 0\big)$ and $\bu_\parallel = \big(0, 0, u_z (x, y)\big)$,
and are periodic in $x$ and $y$.
Two different magnetic Reynolds numbers $Rm_\perp$ and $Rm_\parallel$ are introduced
characterizing the magnitudes of $\bu_\perp$ and $\bu_\parallel$, respectively.
If the flow described by $\bu$ shows helical structures as, e.g., in the case
of the so--called Roberts flow, an $\alpha$-effect occurs such
that $\bscE = \alpha_\perp \, \bmB_\perp$, with some coefficient $\alpha_\perp$,
and $\bmB_\perp$ defined analogously to $\bu_\perp$;
contributions to $\bscE$ with derivatives of $\bmB$ are here ignored.
In the second--order correlation approximation the result for $\alpha_\perp$,
in this case called $\alpha_\perp^{(\mathrm{SOCA})}$,
reads $\alpha_\perp^{(\mathrm{SOCA})} = \alpha_0 Rm_\perp Rm_\parallel$,
with some $\alpha_0$ depending on the geometrical properties but not on the magnitude of the flow.
In the general case it turns out that $\alpha_\perp = \alpha_0 Rm_\perp Rm_\parallel \phi (Rm_\perp)$
with the same $\alpha_0$ as before.
The function $\phi$ satisfies $\phi (0) = 1$, decays monotonically with growing $Rm_\perp$
and tends to zero as $Rm_\perp \to \infty$;
further properties and numerical values of $\phi$ are given in some of the papers mentioned
\citep{raedleretal97,raedleretal02}.
When writing in the sense of (\ref{eq409}) $\alpha_\perp = (1 - f) \alpha_\perp^{(\mathrm{SOCA})}$
we have $f = 1 - \phi$ and therefore $f \to 0$ as $Rm_\perp \to 0$ and $f \to 1$ if $Rm_\perp \to \infty$.

\section{Summary}
\label{sum}

The paper provides a critical view on different analytical ways to determine
the mean electromotive force $\bscE$ in mean--field electrodynamics.
First some of the findings gained with the traditional approach,
reduced to the second--order correlation approximation or approach (i),
are summarized (section~\ref{seci}).
In this context also some new results concerning the case of magnetohydrodynamic background turbulence
are given (section~\ref{subsec32}).
Then the essentials of two versions of the $\tau$--approach, or approach (ii), are represented,
that is, of the simple $\tau$--approach, or approach (iia),
as used by \citet{vainshteinetal83}, \citet{blackmanetal02b} and \citet{brandenburgetal05b},
and of the spectral $\tau$--approach, or approach (iib),
used by \cite{raedleretal03}
and by \citet{rogachevskiietal03,rogachevskiietal04} (section~\ref{secii}).

Approach (i) in its original form, applying to purely hydrodynamic background turbulence,
is based on solutions of the induction equation for the fluctuations $\bbb$
simplified by some approximation.
In the case of magnetohydrodynamic background turbulence in addition solutions of the
momentum balance for $\bu$, again in some approximation, are used.
In the approaches (ii) these equations are not really solved.
Instead, some assumptions on crucial quantities are introduced.
There is hardly any doubt in the correctness of the results of approach (i) in the range of its applicability,
which is well defined at least by sufficient conditions.
It is not surprising that the approaches (ii) deliver the same vectorial structures of the contributions
to $\bscE$ as approach (i), for these are already determined by elementary symmetry principles.
The results of the two types of approaches differ however in the dependence of the coefficients
of the individual contributions to $\bscE$ on the correlation properties
of the velocity fluctuations $\bu$.
The discrepancies of the two approaches as well as strange aspects or shortcomings of the approaches (ii)
are discussed in detail (sections~\ref{subsubseciia2} and \ref{seciibc3}).
Unless any overlap of the ranges of applicability of the approaches (i) and (ii) can be excluded,
what would be very surprising, we have to conclude that the approaches (ii) are in some conflict
with the basic equations and not all of their results can be taken for granted.
For instance, the magnetic contribution to the $\alpha$--effect occurs in the approaches (ii)
not only in the case of a magnetohydrodynamic background turbulence
but for purely hydrodynamic background turbulence, too, where they should not exist.
In addition, in the first case it is the same ``correlation time" which occurs
with the kinetic and the magnetic contributions,
what seems to be in conflict with numerical simulations (section~\ref{subsec43}).
Further the dependence of $\alpha$ and $\beta$
on the magnetic diffusivity $\eta$ and the kinematic viscosity $\nu$ which occurs in approach (iib)
is not correct.
Moreover, it is doubtful whether the approaches (ii) can describe, e.g., the $\bOmega \x \bJ$--effect correctly.

There is no hint that the mentioned shortcomings of the approaches (ii) result from the $\tau$--approximation,
that is, the reduction of third--order correlations of $\bu$ or $\bbb$ to second--order ones.
An important reason for them seems to be an improper treatment of a term connected with the forcing term
in the momentum balance.

A proposal is made for a new approach which avoids any conflict with approach (i) (section~\ref{new}).
In this context also a new formalism for the higher--order correlation approximation is presented
(section~\ref{new1}).

\section*{Acknowledgement}

We thank our referee Eric Blackman for stimulating discussions and numerous helpful critical comments,
likewise Axel Brandenburg for clarifying comments and explanations,
in particular concerning the numerical simulations addressed in section~\ref{subsec43}.
We further thank Igor Rogachevskii and Nathan Kleeorin for many discussions and for drawing
our attention to the ideas described in appendix~\ref{app2}.

\appendix

\section{The limits $q \to \infty$ and $q \to 0$}
\label{app1}

The relations (\ref{eq213}) for $a_{ip}$ and $b_{ipq}$ can be written in the form
\begin{eqnarray}
a_{ip} &=&  \frac{\epsilon_{ijn} \delta_{lp} - \epsilon_{ijp} \delta_{ln}}{2 (4 \pi)^{3/2}} \,
     q^{5/2} \frac{\tau_c}{\lambda_c} \,
     \int_\infty \int_{0}^\infty
     \frac{\exp(- q {\xi'}^2 / 4 \tau')}{{\tau'}^{5/2}} \,
     Q_{jn} (\bx, t; \bxi' \lambda_c, - \tau' \tau_c) \, {\xi'}_l \, \dd^3 \xi' \, \dd \tau'
\nonumber\\
&=&  \frac{\epsilon_{ijn} \delta_{lp} - \epsilon_{ijp} \delta_{ln}}{(4 \pi)^{3/2}} \,
     q^{3/2} \frac{\tau_c}{\lambda_c} \,
     \int_\infty \int_{0}^\infty
     \frac{\exp(- q {\xi'}^2 / 4 \tau')}{{\tau'}^{3/2}} \,
     \frac{\p Q_{jn} (\bx, t; \bxi' \lambda_c, - \tau' \tau_c)}{\p {\xi'}_l} \, \dd^3 \xi' \, \dd \tau'
\nonumber\\
b_{ipq} &=& \frac{\epsilon_{ijn} \delta_{lp} - \epsilon_{ijp} \delta_{ln}}{2 (4 \pi)^{3/2}}
\label{eq223}\\
&& \quad \quad
     q^{5/2} \, \tau_c \, \int_\infty \int_{0}^\infty
     \frac{\exp(- q {\xi'}^2 / 4 \tau')}{{\tau'}^{5/2}} \,
     Q_{jn} (\bx, t; \bxi' \lambda_c, - \tau' \tau_c) \, {\xi'}_l {\xi'}_q \, \dd^3 \xi' \, \dd \tau' \, .
\nonumber
\end{eqnarray}

In order to evaluate (\ref{eq223}) in the limit $q \to \infty$
it is useful to introduce the new integration variable $\xi'' = \sqrt{q} \xi'$.
Then $q$ occurs at no other places in these relations than in the third argument of $Q_{jn}$,
which takes the form $\xi'' \lambda_c / \sqrt{q}$.
Note that $\xi'' \lambda_c / \sqrt{q} = \xi'' \sqrt{\eta \tau_c}$.
We may have $q \to \infty$ as a consequence of $\eta \tau_c \to 0$ or of $\lambda_c \to \infty$.
In the first case this third argument tends to zero as $q \to \infty$.
In the second case, which seems rather academic, $Q_{jn}$ looses its dependence on this argument as $q \to \infty$,
and we may replace the latter simply by zero.
Considering then
\begin{equation}
\int_\infty \frac{\exp(- \xi^2 / 4 \tau)}{\tau^{3/2}} \, \dd^3 \xi
     = (4 \pi)^{3/2} \, , \quad
\int_\infty \frac{\exp(- \xi^2 / 4 \tau)}{\tau^{5/2}} \, \xi_l \, \xi_q \, \dd^3 \xi
     = 2 \,(4 \pi)^{3/2} \, \delta_{lq}
\label{eq225}
\end{equation}
we find (\ref{eq227}).

In view of the limit $q \to 0$ we introduce $\tau'' = \tau' / q$.
Then $q$ occurs only in the fourth argument of $Q_{jn}$,
that is, $\tau'' q  \tau_c = \tau'' \lambda^2_c / \eta$.
We may have $q \to 0$ due to $\lambda^2_c / \eta \to 0$ or due to $\tau_c \to \infty$.
In the first case the fourth argument tends to zero as $q \to 0$.
In the second one $Q_{jn}$ looses its dependence on this argument as $q \to 0$,
and we may replace it by zero.
With
\begin{equation}
\int_0^\infty \frac{\exp(- \xi^2 / 4 \tau)}{\tau^{3/2}} \, \dd \tau
    = \frac{2 \sqrt{\pi}}{\xi} \, , \quad
\int_0^\infty \frac{\exp(- \xi^2 / 4 \tau)}{\tau^{5/2}} \, \dd \tau
    = \frac{4 \sqrt{\pi}}{\xi^3}
\label{eq229}
\end{equation}
we find then (\ref{eq231}).

\section{An alternative representation of $\alpha$ and $\beta$ as given by (\ref{eq241})}
\label{app2}

Let us consider a homogeneous statistically steady turbulence.
In this case the correlation tensor $Q_{ij}$ defined by (\ref{eq208}) depends no longer
on $\bx$ and $t$, that is $Q_{ij} = Q_{ij} (\bxi, \tau)$.
We introduce the quantity $U_{ij}$ by
\begin{equation}
U_{ij} = \int_0^\infty  Q_{ij} (\bzo, - \tau) \, \dd \tau \, .
\label{eq503}
\end{equation}
With the Fourier transform ${\hat{Q}}_{ij}$ of $Q_{ij}$ it holds
\begin{equation}
U_{ij} = \int \int_0^\infty  {\hat{Q}}_{ij} (\bk, - \tau) \, \dd^3 k \, \dd \tau \, .
\label{eq505}
\end{equation}

Assume now that ${\hat{Q}}_{ij}$ allows a separation ansatz of the form
\begin{equation}
{\hat{Q}}_{ij}  (\bk, \tau) = {\hat{Q}}_{ij}  (\bk, 0) \, f (k, \tau) \, ,
\label{eq507}
\end{equation}
where, of course, $f (k, 0) = 1$.
Then we have
\begin{equation}
U_{ij} = \int  {\hat{Q}}_{ij} (\bk, 0) {\hat{\tau}}_c (k) \, \dd^3k
      = (2 \pi)^{-3} \int Q_{ij} (\bxi, 0) \, \tau_c (\xi) \, \dd^3\xi \, ,
\label{eq509}
\end{equation}
where ${\hat{\tau}}_c (k)$ is defined by
\begin{equation}
{\hat{\tau}}_c (k) = \int_0^\infty f (k, - \tau) \, \dd \tau \, ,
\label{eq511}
\end{equation}
and $\tau_c (\xi)$ such that ${\hat{\tau}}_c$ is its Fourier transform,
\begin{equation}
\tau_c (\xi) = \int {\hat{\tau}}_c (k) \, \exp( \iu \bk \cdot \bxi ) \, \dd^3 k \, .
\label{eq513}
\end{equation}
Comparing (\ref{eq503}) and (\ref{eq509}) and remembering (\ref{eq208})
we arrive then at the remarkable relation
\begin{equation}
\int_0^\infty  \langle u_i (\bx, t) \, u_j (\bx, t - \tau) \rangle \, \dd \tau
     = (2 \pi)^{-3} \int \langle u_i (\bx, t) \, u_j (\bx + \bxi, t) \rangle \, \tau_c (\xi) \, \dd^3\xi \, .
\label{eq515}
\end{equation}

We restrict or attention now to an incompressible fluid and assume the turbulence to be isotropic.
As is well known the Fourier transform ${\hat{Q}}_{ij}$ of $Q_{ij}$ has then the form
\begin{equation}
{\hat{Q}}_{ij} (\bk, \tau) = \big( \delta_{ij} - \frac{k_i k_j}{k^2} \big) \, g (k, \tau)
     + \iu \, \epsilon_{ijk} \, \frac{k_k}{k^2} \, h (k, \tau) \, .
\label{eq517}
\end{equation}
For the calculation of $\beta$ according to (\ref{eq237}) it is sufficient
to consider a mirror--symmetric turbulence, that is $h = 0$.
Then ${\hat{Q}}_{ij}$ satisfies the assumption (\ref{eq507}).
Consequently (\ref{eq515}) applies, and $\beta$ as given by (\ref{eq241})
can also be written in the form
\begin{equation}
\beta = \frac{1}{3 (2 \pi)^3}
     \int \langle \bu (\bx, t) \cdot \bu (\bx + \bxi, t) \rangle \, \tau_c (\xi) \, \dd^3\xi \, .
\label{eq519}
\end{equation}

We may repeat the above derivations with
\begin{equation}
{\tilde{Q}}_{ij} (\bxi, \tau) = \epsilon_{jkl} \frac{\p Q_{il} (\bxi, \tau)}{\p \xi_k}
\label{eq521}
\end{equation}
instead of $Q_{ij} (\bxi, \tau)$.
Using again (\ref{eq517}) we have then
\begin{equation}
{\hat{\tilde{Q}}}_{ij} (\bk, \tau) = \big( \delta_{ij} - \frac{k_i k_j}{k^2} \big) \, h (k, \tau)
     + \iu \epsilon_{ijk} k_k \, g (k, \tau)  \, .
\label{eq522}
\end{equation}
For the calculation of $\alpha$ it seems to be justified to put here $g = 0$.
Then (\ref{eq522}) satisfies the analogue to (\ref{eq507}), and we find an analogue to (\ref{eq515})
with $u_j$ replaced with $(\bnab \x \bu)_j$.
With this result $\alpha$ as given by (\ref{eq241})
can be written in the form
\begin{equation}
\alpha = - \frac{1}{3 (2 \pi)^3}
     \int \langle \bu (\bx, t) \cdot \big(\bnab \x \bu (\bx + \bxi, t)\big) \rangle \, \tau_c (\xi) \, \dd^3\xi \, .
\label{eq523}
\end{equation}
The quantities $\tau_c$ in (\ref{eq519}) and (\ref{eq523}) may be well different from each other.

\section{$\bOmega \x \bJ$--effect}
\label{app3}
Under the assumptions which justify equation (\ref{eq115}) we have
\begin{equation}
b_{ijk} = \beta \epsilon_{ijk} - \delta_1 \delta_{ij} \, \Omega_k - \delta_2 \delta_{ik} \, \Omega_j
    - \delta_3 \delta_{jk} \, \Omega_i \, .
\label{eq801}
\end{equation}
The $\beta$, $\delta_1$ and $\delta_2$ terms lead just to the last line of (\ref{eq115})
whereas the $\delta_3$ term has because of $\bnab \cdot \bmB = 0$ no counterpart there.
Multiplying both sides of (\ref{eq801}) first by $\delta_{ij} \, \Omega_k$ and then by $\delta_{ik} \, \Omega_j$
and by $\delta_{jk} \, \Omega_i$ we arrive at a system of three equations
for $\delta_1$, $\delta_2$ and $\delta_3$.
Its solutions for $\delta_1$ and $\delta_2$ read
\begin{equation}
\delta_1 = - \frac{1}{10 \Omega^2} \, (4 b_{iij} - b_{iji} - b_{jii}) \, \Omega_j \, , \quad
\delta_2 = \frac{1}{10 \Omega^2} \, (b_{iij} - 4 b_{iji} + b_{jii}) \, \Omega_j \, .
\label{eq803}
\end{equation}
Expressing $b_{ijk}$ according to (\ref{eq213}) we see
that $\delta_1$ depends only on the part of $Q_{ij}$ which is firstly antisymmetric in the indices
and secondly even in $\bxi$.
Assuming that the turbulence is homogeneous and statistically steady,
which implies $Q_{ij} (\bxi, \tau) = Q_{ji} (-\bxi, -\tau)$, we may conclude that this part is
odd in $\tau$ and therefore vanishes at $\tau = 0$.
As for $\delta_2$ this statement applies only under the additional assumption of an incompressible fluid,
that is, $\p Q_{ij} / \p \xi_j = 0$.

These conclusions can also be drawn immediately from the first results
on the $\bOmega \x \bJ$--effect given by \citet{raedler69}.
By the reason explained above, the functions $h$ and $k$ used there are odd
but $l$ is even in $\tau$, and it can be easily shown that $l$ vanishes in the incompressible case.
This just confirms the above statements on the part of $Q_{ij}$
responsible for the coefficients that correspond to $\delta_1$ and $\delta_2$.

By the way, the reason why the $\bOmega \x \bJ$--effect did not occur in the first calculation
of the mean electromotive force $\bscE$ delivered by \citet{steenbecketal66}
was a not completely correct ansatz for $Q_{ij}$, just without a part being odd in $\tau$.

\section{Derivation of equation (\ref{eq337})}
\label{app4}

We start from (\ref{eq333}) and express $v_i$ and $w_j$ on the right--hand side
by their Fourier representations,
\begin{equation}
\Phi_{ij} (\bv, \bw; \bx, \bxi) = \int \int \langle {\hat{v}}_i (\bk') \, {\hat{w}}_j (\bk'') \rangle \,
     \exp\big( \, \iu \big( \bk' \cdot (\bx + \bxi/2) + \bk'' \cdot (\bx - \bxi/2) \big ) \big) \,
     \dd^3 k' \, \dd^3 k'' \, .
\label{eq601}
\end{equation}
Introducing new integration variables $\bk$ and $\bK$
by $\bk' = \bk + \bK/2$ and $\bk'' = - \bk + \bK/2$ we find then
\begin{eqnarray}
\Phi_{ij} (\bv, \bw; \bx, \bxi) &=& \int {\hat{\Phi}}_{ij} (\bx, \bk) \exp( \iu \bk \cdot \bxi ) \, \dd^3 k
\nonumber\\
{\hat{\Phi}}_{ij} (\bv, \bw; \bx, \bk) &=& \int \langle {\hat{v}}_i (\bk + \bK/2) \,
    {\hat{w}}_j (-\bk + \bK/2) \rangle \, \exp( \iu \bK \cdot \bx ) \, \dd^3 K \, .
\label{eq603}
\end{eqnarray}
The last line is identical to (\ref{eq337}).

\section{Higher--order correlation approximations}
\label{app5}

Consider again an infinitely extended homogeneous fluid and assume for the sake of simplicity
$\bmU = \bzo$ and $\bnab \cdot \bu = 0$.
Assume in view of the calculation of $\bscE$ at $\bx = \tilde{\bx}$ that
\begin{equation}
\mB_i = B_i + B_{ij} (x_j - {\tilde{x}}_j) \, ,
\label{eq701}
\end{equation}
with $B_i$ and $B_{ij}$ independent of position and steady, and $B_{ii} = 0$.
Clearly $B_i$ and $B_{ij}$ are equal to $\mB_i$ and $\p \mB_i / \p x_j$, respectively, at $\bx = \tilde{\bx}$.
From (\ref{eq107}) and (\ref{eq109}) it follows that
\begin{equation}
\p_t b_k - \eta \Delta b_k = \big(B_p + B_{pq} (x_q - {\tilde{x}}_q)\big) \, \frac{\p u_k}{\p x_p} - B_{kp} u_p
    + (\delta_{kq} \delta_{pr} - \delta_{kr} \delta_{pq}) \frac{\p}{\p x_p} \big( u_q b_r \big)' \, , \quad
\frac{\p b_i}{\p x_i} = 0
\label{eq703}
\end{equation}
and, if again the influence of the initial $\bbb$ is ignored, that is $t_0 \to - \infty$,
\begin{eqnarray}
b_k (\bx, t) &=& B_p \, \int_0^\infty \!\!\! \int_\infty
    \frac{\p G^{(\eta)} (\xi, \tau)}{\p \xi_p} \, u_k (\bx - \bxi, t - \tau) \, \dd^3 \xi \dd \tau
\nonumber\\
&& \!\!\!\!\!\!\!\!\!\!\!\!\!\!\! - \, B_{pq} \, \int_0^\infty \!\!\! \int_\infty
    \Big( G^{(\eta)} (\xi, \tau) \, \delta_{kp} \, \delta_{qr} -
    \frac{\p G^{(\eta)} (\xi, \tau)}{\p \xi_p} \, (x_q - {\tilde{x}}_q - \xi_q) \, \delta_{kr} \Big) \,
    u_r (\bx - \bxi, t - \tau) \, \dd^3 \xi \dd \tau
\label{eq705}\\
&& + (\delta_{kq} \delta_{pr} - \delta_{kr} \delta_{pq}) \, \int_0^\infty \!\!\! \int_\infty
    \frac{\p G^{(\eta)} (\xi, \tau)}{\p \xi_p} \, \big( u_q (\bx - \bxi, t - \tau) \,
    b_r (\bx - \bxi, t - \tau) \big)' \, \dd^3 \xi \dd \tau \, .
\nonumber
\end{eqnarray}

Put then
\begin{equation}
b_k = b^{(1)}_k + b^{(2)}_k + b^{(3)}_k + \ldots
\label{eq707}
\end{equation}
and obtain
\begin{eqnarray}
b^{(1)}_k (\bx, t) &=& B_p \, \int_0^\infty \!\!\! \int_\infty
    \frac{\p G^{(\eta)} (\xi, \tau)}{\p \xi_p} \, u_k (\bx - \bxi, t - \tau) \, \dd^3 \xi \dd \tau
\nonumber\\
&& \!\!\!\!\!\!\!\!\!\!\!\!\!\!\! - \, B_{pq} \, \int_0^\infty \!\!\! \int_\infty
    \Big( G^{(\eta)} (\xi, \tau) \, \delta_{kp} \, \delta_{qr} -
    \frac{\p G^{(\eta)} (\xi, \tau)}{\p \xi_p} \, (x_q - {\tilde{x}}_q - \xi_q) \, \delta_{kr} \Big) \,
    u_r (\bx - \bxi, t - \tau) \, \dd^3 \xi \dd \tau
\label{eq709}\\
b_k^{(\nu)} (\bx, t) &=&  (\delta_{kq} \delta_{pr} - \delta_{kr} \delta_{pq}) \,
    \int_0^\infty \!\!\! \int_\infty
    \frac{\p G^{(\eta)} (\xi, \tau)}{\p \xi_p} \big( u_q (\bx - \bxi, t - \tau) \,
    b^{(\nu - 1)}_r (\bx - \bxi, t - \tau) \big)' \, \dd^3 \xi \, \dd \tau
\nonumber\\
&& \qquad \qquad \qquad  \qquad \qquad \qquad  \qquad \qquad \qquad \qquad \qquad \qquad \qquad \qquad
    \nu \geq 2 \, .
\nonumber
\end{eqnarray}
These relations are equivalent to
\begin{eqnarray}
b^{(\nu)}_k (\bx, t) &=& B_p \, \int_0^\infty \!\!\! \int_\infty
    \frac{\p G^{(\eta)} (\xi, \tau)}{\p \xi_p} \big( v^{(\nu)}_k (\bx, t; - \bxi, - \tau) \big)' \,
    \dd^3 \xi \, \dd \tau
\nonumber\\
&& + B_{pq} \, \int_0^\infty \!\!\! \int_\infty
    G^{(\eta)} (\xi, \tau) \, \big( w^{(\nu)}_{kpq} (\bx, t; - \bxi, - \tau) \big)' \,
    \dd^3 \xi \, \dd \tau
    \, , \quad \nu \geq 1 \, ,
\label{eq711}
\end{eqnarray}
with
\begin{eqnarray}
v^{(1)}_k (\bx, t; \bxi, \tau) &=& u_k (\bx + \bxi, t + \tau) \, , \quad
\nonumber\\
v^{(\nu)}_k (\bx, t; \bxi, \tau) &=&
   \int_0^\infty \!\!\! \int_\infty \frac{\p G^{(\eta)} (\xi', \tau')}{\p \xi'_l} \,
   \Big(u_k (\bx - \bxi', \tau - \tau') \, v^{(\nu-1)}_l (\bx - \bxi', t - \tau'; \bxi, \tau)
\label{eq713}\\
&& \qquad \qquad
   - u_l (\bx - \bxi', \tau - \tau') \, v^{(\nu-1)}_k (\bx - \bxi', t - \tau'; \bxi, \tau) \Big) \,
   \dd^3 \xi' \dd \tau' \, , \quad \nu \geq 2 \, ,
\nonumber
\end{eqnarray}
and
\begin{eqnarray}
w^{(1)}_{kpq} (\bx, t; \bxi, \tau) &=& - \, \delta_{kp} u_q (\bx + \bxi, t + \tau)
   + (x_q - {\tilde{x}}_q + \xi_q) \frac{\p u_k (\bx + \bxi, t + \tau)}{\p \xi_p}
\nonumber\\
w^{(\nu)}_k (\bx, t; \bxi, \tau) &=&
   \int_0^\infty \!\!\! \int_\infty \frac{\p G^{(\eta)} (\xi', \tau')}{\p \xi'_l} \,
   \Big(u_k (\bx - \bxi', \tau - \tau') \, w^{(\nu-1)}_{lpq} (\bx - \bxi', t - \tau'; \bxi, \tau)
\label{eq715}\\
&& \qquad \qquad
   - u_l (\bx - \bxi', \tau - \tau') \, w^{(\nu-1)}_{kpq} (\bx - \bxi', t - \tau'; \bxi, \tau) \Big)
   \dd^3 \xi' \dd \tau' \, , \quad \nu \geq 2 \, .
\nonumber
\end{eqnarray}

From (\ref{eq707}) and (\ref{eq711}) it can be concluded that
\begin{eqnarray}
b_k (\bx, t) &=& B_p \, \int_0^\infty \!\!\! \int_\infty
    \frac{\p G^{(\eta)} (\xi, \tau)}{\p \xi_p} \big( v_k (\bx, t; - \bxi, - \tau) \big)' \,
    \dd^3 \xi \, \dd \tau
\nonumber\\
&& + B_{pq} \, \int_0^\infty \!\!\! \int_\infty
    G^{(\eta)} (\xi, \tau) \, \big( w_{kpq} (\bx, t; - \bxi, - \tau) \big)' \,
    \dd^3 \xi \, \dd \tau \, ,
\label{eq717}
\end{eqnarray}
where
\begin{eqnarray}
v_k (\bx, t; \bxi, \tau) &=&  v^{(1)}_k (\bx, t; \bxi, \tau) + v^{(2)}_k (\bx, t; \bxi, \tau)
    + v^{(3)}_k (\bx, t; \bxi, \tau) + \ldots
\nonumber\\
w_{kpq} (\bx, t; \bxi, \tau) &=&  w^{(1)}_{kpq} (\bx, t; \bxi, \tau) + w^{(2)}_{kpq} (\bx, t; \bxi, \tau)
    + w^{(3)}_{kpq} (\bx, t; \bxi, \tau) + \ldots \, ,
\label{eq719}
\end{eqnarray}
with $v^{(\nu)}_k$ and $w^{(\nu)}_{kpq}$ as given by (\ref{eq713}) and (\ref{eq715}).

Calculating then ${\cE}_i = \epsilon_{ijk} \langle u_j b_k \rangle$ and expanding it according to (\ref{eq111})
we finally find
\begin{eqnarray}
a_{ip} (\tilde{\bx}, t) &=&  \epsilon_{ijk}
   \int_0^\infty \!\!\! \int_\infty  \frac{\p G^{(\eta)} (\xi, \tau)}{\p \xi_p} \,
   \ol{u_j (\tilde{\bx}, t) \, v_k (\tilde{\bx}, t; - \bxi, - \tau)} \, \dd^3 \xi \, \dd \tau \, \quad
\nonumber\\
b_{ipq} (\tilde{\bx}, t) &=&  \epsilon_{ijk}
   \int_0^\infty \!\!\! \int_\infty  G^{(\eta)} (\xi, \tau) \,
   \ol{u_j (\tilde{\bx}, t) \, w_{kpq} (\tilde{\bx}, t; - \bxi, - \tau)} \, \dd^3 \xi \, \dd \tau \, ,
\label{eq721}
\end{eqnarray}
which has to be completed by (\ref{eq713}), (\ref{eq715}) and (\ref{eq719}).

\newpage

\noindent
{\bf Note added in proof}

\bigskip

\noindent In appendix~\ref{app2} it has been shown that the results (\ref{eq241})
of approach (i) for $\alpha$ and $\beta$, with integrals
of $\langle \bu (\bx, t) \cdot (\bnab \x \bu (\bx, t - \tau)) \rangle$
and $\langle \bu (\bx, t) \cdot \bu (\bx, t - \tau) \rangle$ over $\tau$,
can be rewritten into (\ref{eq519}) and (\ref{eq523}), with integrals containing $\bu$
only as $\langle \bu (\bx, t) \cdot (\bnab \x \bu (\bx + \bxi, t)) \rangle$
or $\langle \bu (\bx, t) \cdot \bu (\bx + \bxi, t) \rangle$, taken over all $\bxi$.
We have argued there with very specific properties of the turbulence, that is of $\bu$.

Relations between the two types of such integrals, over $\tau$ and over $\bxi$,
exist also in a much wider range of assumptions.
Consider any function $F = F (\bxi, \tau)$.
Assume that the integral $\int_0^\infty F (\bzo, - \tau) \, \dd \tau$ has a non-zero finite value.
Assume further that the integral $\int_\infty F (\bxi, 0) \, g (\bxi) \, \dd^3 \xi$,
with any weight function $g$, has a non--zero finite value, too.
Clearly the magnitude of a given $g$ can be fixed such that
\[
\qquad \qquad \qquad \qquad \qquad \quad
\int_0^\infty F (\bzo, - \tau) \, \dd \tau = (2 \pi)^{-3} \, \int_\infty F (\bxi, 0) \, g (\bxi) \, \dd^3 \xi \, .
\qquad \qquad \qquad \qquad \qquad \qquad \quad (\star)
\]
There is, of course, some freedom in the choice of the dependence of $g$ on $\bxi$.
We may determine a specific $g$, e.g., after the pattern of appendix~\ref{app2}.
For this purpose we introduce the Fourier transform $\hat{F}$ of $F$ with respect to $\bxi$
and put $\int_0^\infty \hat{F} (\bk, - \tau) \, \dd \tau = \hat{F} (\bk, 0) \, \hat{h} (-\bk)$.
Integrating both sides of this relation over all $\bk$
and using the convolution theorem we arrive just at ($\star$),
with $g (\bxi)$ such that its Fourier transform is equal to $\hat{h} (\bk)$.

Let us first identify $F$ in relation ($\star$)
with $\langle \bu (\bx, t) \cdot (\bnab \x \bu (\bx + \bxi, t - \tau)) \rangle$
or $\langle \bu (\bx, t) \cdot \bu (\bx + \bxi, t - \tau) \rangle$.
Then we arrive without utilizing specific properties of the turbulence
at relations which allow us to rewrite the results (\ref{eq241}) for $\alpha$ and $\beta$
into (\ref{eq519}) and (\ref{eq523}).
In the same way we may rewrite the more general results (\ref{eq227}) for $a_{ip}$ and $b_{ipq}$
such that, as in (\ref{eq381}), only correlations of components of $\bu$ at the same time
and only integrals over $\bxi$ occur.

This finding extenuates our statement made in section~\ref{seciibc3}
on the discrepancy between the results (\ref{eq227}) of approach (i)
and the results (\ref{eq381}) of approach (iib).
However, it probably not even completely resolves the part of the discrepancy
resulting from the different types of correlations and different types of integrals:
So it is hard to imagine that all integrals in (\ref{eq227}) over correlations of $\bu$
can be rewritten as integrals over $\bxi$ with one and the same $g$,
which would have then to be identified with the $\tau_*$ that occurs in (\ref{eq381}).

\end{document}